\documentclass[lettersize,journal]{IEEEtran}
\usepackage{amsmath,amsfonts}
\usepackage{array}
\usepackage{textcomp}
\usepackage{stfloats}
\usepackage{verbatim}
\usepackage{cite}
\usepackage{amsmath}
\usepackage{comment}
\usepackage{multirow}
\usepackage[ruled]{algorithm2e}
\usepackage{graphicx}
\usepackage{lipsum}
\usepackage{url}

\usepackage{subfigure}
\usepackage{caption}
\usepackage{booktabs,chemformula}
\usepackage{rotating}
\usepackage{fancyhdr}
\usepackage{gensymb}
\usepackage{enumitem}
\usepackage{color}
\usepackage{xcolor,colortbl}

\usepackage{algpseudocode}
\usepackage{algcompatible}
\definecolor{Gray}{gray}{0.85}
\usepackage{blindtext} 
\usepackage{subfiles}
\usepackage[T1]{fontenc}
\usepackage[utf8]{inputenc}
\begin{document}

\title{Analysis of Loss and Crosstalk Noise in MZI-based Coherent Silicon Photonic Neural Networks}

\author{Amin Shafiee$^1$,~\IEEEmembership{Student Member,~IEEE}, Sanmitra Banerjee$^2$,~\IEEEmembership{Student Member,~IEEE}, Krishnendu Chakrabarty$^3$,~\IEEEmembership{Fellow,~IEEE}, Sudeep Pasricha$^1$,~\IEEEmembership{Senior Member,~IEEE}, and Mahdi Nikdast$^1$,~\IEEEmembership{Senior Member,~IEEE} 
$^1$Department of Electrical and Computer Engineering, Colorado State University, Fort Collins, CO 80523, USA\\
$^2$NVIDIA, Santa Clara, CA 95134, USA \\
$^3$ Department of Electrical, Computer and Energy Engineering, Arizona State University, Tempe, AZ 85281, USA \\
\thanks{This work was supported in part by the National Science Foundation (NSF) under grant numbers CCF-2006788 and CNS-2046226.}
}
\maketitle

\begin{abstract}

With the continuous increase in the size and complexity of machine learning models, the need for specialized hardware to efficiently run such models is rapidly growing. To address such a need, silicon-photonic-based neural network (SP-NN) accelerators have recently emerged as a promising alternative to electronic accelerators due to their lower latency and higher energy efficiency. Not only can SP-NNs alleviate the fan-in and fan-out problem with linear algebra processors, their operational bandwidth can match that of the photodetection rate (typically $\approx$100 GHz), which is at least over an order of magnitude faster than electronic counterparts that are restricted to a clock rate of a few GHz. Unfortunately, the underlying silicon photonic devices in SP-NNs suffer from inherent optical losses and crosstalk noise originating from fabrication imperfections and undesired optical couplings, the impact of which accumulates as the network scales up. Consequently, the inferencing accuracy in an SP-NN can be affected by such inefficiencies---e.g., can drop to below 10\%---the impact of which is yet to be fully studied. In this paper, we comprehensively model the optical loss and crosstalk noise using a bottom-up approach, from the device to the system level, in coherent SP-NNs built using Mach--Zehnder interferometer (MZI) devices. The proposed models can be applied to any SP-NN architecture with different configurations to analyze the effect of loss and crosstalk. Such an analysis is important where there are inferencing accuracy and scalability requirements to meet when designing an SP-NN. Using the proposed analytical framework, we show a high power penalty and a catastrophic inferencing accuracy drop of up to 84\% for SP-NNs of different scales with three known MZI mesh configurations (i.e., Reck, Clements, and Diamond) due to accumulated optical loss and crosstalk noise.
\end{abstract}


\section{Introduction} 
\label{sec:: 1-Introduction}
With the rising demand for larger neural networks to address complex and computationally expensive problems, artificial intelligence (AI) accelerators need to consistently deliver better performance and improved accuracy while being energy-efficient. In this context, deep neural networks have attracted substantial interest for various applications ranging from image recognition to network anomaly detection, decision-making problems, self-driving cars, pandemic rate prediction, and early-stage cancer detection \cite{ONN_survey}. Given the significant growth in the demand for data-driven and computationally expensive applications, the energy efficiency of electronic-based deep-learning inference accelerators has been relatively low, and they have been unable to keep up with the performance requirements of emerging deep-learning applications \cite{SiPh_codesign, pasricha2020survey}.
\begin{figure}[t]
\centering
\includegraphics[width=3.5in]{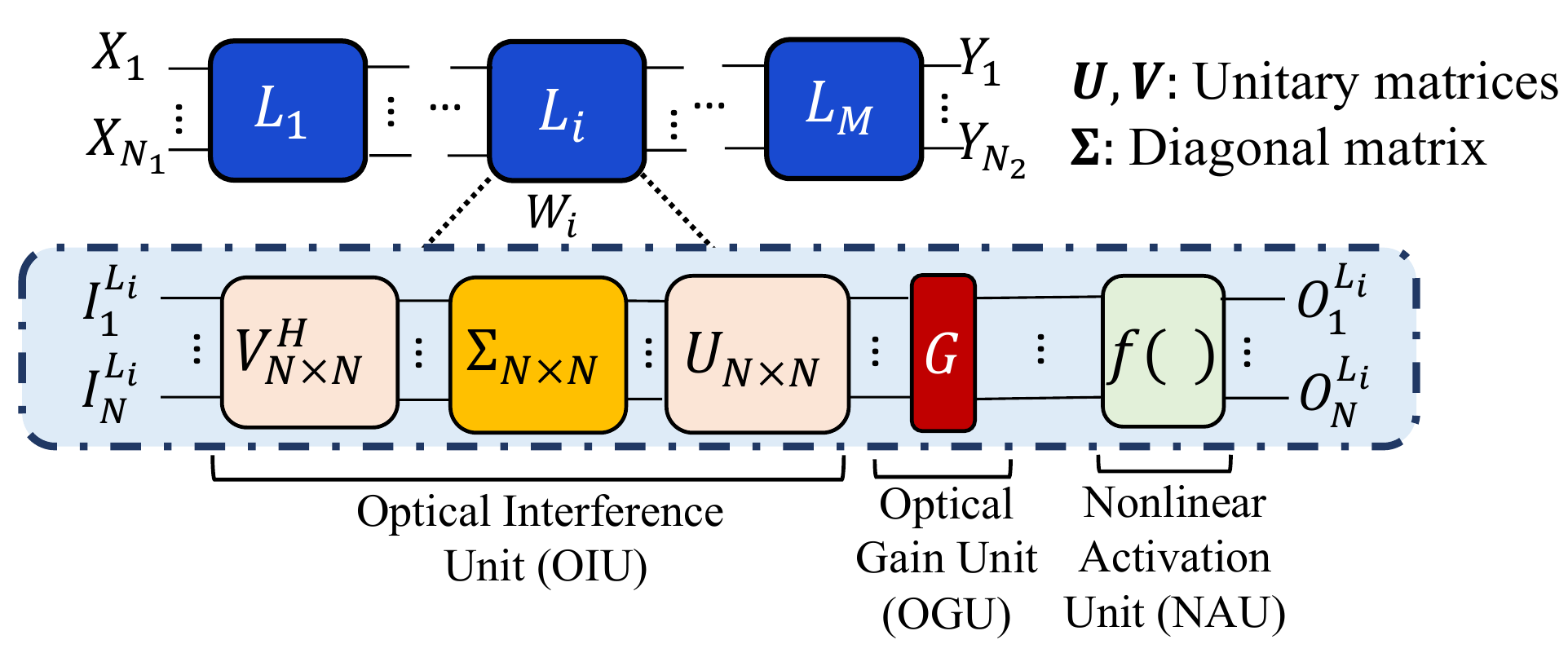}
  \caption{ Overview of a coherent SP-NN with $N_1$ inputs, $N_2$ outputs, and $M$ layers. }\label{Fig:: MZI-Structure}
  \vspace{-0.15in}
\end{figure}
Silicon photonics (SiPh) enables the deployment of integrated photonics across a wide range of applications, from realizing ultra-fast communication for Datacom applications \cite{palmieri2020enhanced, 10088445_yogesh, hamdani2021highly, verma2022temperature} to energy-efficient optical computation in emerging hardware accelerators for deep learning \cite{mirza2022characterization,zhang2021optical,wiecha2021deep}. To alleviate the limitations of conventional CMOS-based electronic accelerators in terms of energy consumption and latency, new SiPh-based hardware accelerators optimized for deep learning applications are on the rise. By leveraging optical interconnects for communication and photonic devices for computation, silicon-photonic-based neural network (SP-NN) accelerators offer the promise of up to 1000 times higher energy efficiency for performing computationally expensive multiply-and-accumulate operations \cite{SiPh_codesign}, which are the most power-hungry and common operations in deep learning applications \cite{SiPh_codesign}. 

Among different SP-NN implementations, coherent SP-NNs, which operate on a single wavelength, have an inherent advantage over noncoherent SP-NNs that require power-hungry wavelength-conversion steps and multiple wavelength sources \cite{sunny2021survey}. Fig.~\ref{Fig:: MZI-Structure} presents an overview of a multi-layer coherent SP-NN with $N_1$ inputs, $N_2$ outputs, and $M$ layers. Each layer comprises an optical-interference unit (OIU) implemented using an array of Mach--Zehnder interferometers (MZIs) with a specific architecture, connected to a nonlinear-activation unit (NAU) using an optical gain (amplification) unit (OGU). Within an OIU, MZIs can be used to realize matrix-vector multiplication as shown by \cite{shafiee2022loci, SiPh_codesign}. Accordingly, several coherent SP-NN architectures have been proposed by cascading arrays of MZIs to perform large-scale linear multiplication in the optical domain \cite{shokraneh2020diamond,Clements:16,reck1994experimental}.


While SP-NNs are promising alternatives to electronic-based deep learning hardware accelerators, several factors limit their performance and scalability. The underlying devices in SP-NNs (e.g., MZIs in coherent SP-NNs) suffer from optical loss and crosstalk noise due to device fabrication imperfections (e.g., sidewall roughness) and undesired mode couplings \cite{mirza2022characterization, Bahadori:16}. For example, prior work has shown up to 1.5~dB insertion loss and $-$18~dB crosstalk in a 2$\times$2 MZI \cite{Farhad_4by4}. While optical loss and crosstalk noise are small and seem to be negligible at the device level, they accumulate as the network scales up, hence leading to severe performance degradation at the network and system level (e.g., drop in inferencing accuracy). Moreover, crosstalk cannot be filtered in coherent SP-NNs---our focus in this paper---due to the coherence between the noise and victim signals. Therefore, there is a critical need for careful analysis of optical loss and crosstalk noise in coherent SP-NNs and exploring their impact on SP-NN performance. 

The novel contribution of this paper is to comprehensively analyze optical loss and crosstalk noise and their impact on coherent SP-NN performance from the device level to the system level. We develop a realistic device-level MZI compact model to analyze the optical loss from different sources (i.e., propagation loss, directional coupler loss, and metal absorption loss) and the coherent crosstalk noise in the MZI. This model is also able to capture the impact of optical phase settings, which represent weight parameters in coherent SP-NNs, on the MZI's optical loss and crosstalk noise. Leveraging our accurate device-level models, we present layer- and network-level optical loss and coherent crosstalk models that scale with the number of inputs and layers in coherent SP-NNs. In addition, we propose a detailed analysis of the effect of optical loss and crosstalk in SP-NNs when the optoelectronic NAU units are used. The proposed framework enables an accurate exploration of the laser power penalty and inferencing accuracy drop in SP-NNs with different mesh configurations under the effect of optical loss and crosstalk noise. Leveraging our proposed framework, we also quantify the maximum optical loss acceptable in the underlying devices when specific inferencing accuracy goals must be met within an SP-NN.

The proposed analytical framework can be applied to any coherent SP-NN architecture of any size to analyze the average and worst-case optical loss and crosstalk noise in the network. In this paper, we consider three well-known MZI-based coherent SP-NN architectures, namely Clements \cite{Clements:16}, Reck \cite{reck1994experimental}, and Diamond \cite{shokraneh2020diamond}. For example, for the case study of the Clements SP-NN with 16 inputs and 2 hidden layers, we show that the optical loss, optical crosstalk power, and laser power penalty (i.e., to compensate for optical loss and crosstalk) can be as high as 6~dB, 31.6~dBm, and 20~dBm, respectively. Considering the MNIST classification task as an example, we show that the network inferencing accuracy can drop by about 84.6\% due to optical loss and crosstalk noise. We also show that by increasing the number of inputs from 16 to 64 in the same network, the resulting optical power penalty increases significantly to as high as 140~dBm. The proposed analyses in this paper extend our prior work in \cite{shafiee2022loci} by performing the loss and crosstalk analysis for two more SP-NN configurations, analyzing the effect of optical loss and crosstalk in SP-NNs when optoelectronic nonlinear activation units are used, presenting an optical power penalty model for SP-NNs, and analysing the scalability constraints due to the optical loss and crosstalk in the MZI-based SP-NNs when being used as a photonic processing unit for high-performance computation. 

The rest of the paper is organized as follows. Section \ref{background} presents an overview of the building blocks in SP-NNs, SP-NN design and working mechanism, and prior related work. Section \ref{Loss_XT_framework} presents analytical models to analyze the impact of optical loss and crosstalk from the device level to the system level in SP-NNs. The impact of the optical loss and crosstalk in optoelectronic NAU units is modeled in this section. Section \ref{Simulation} presents the simulation results to show the impact of loss and crosstalk on the performance of SP-NNs with the three MZI mesh configurations of Clements, Reck, and Diamond. Section \ref{power_analysis} presents the discussion on the effect of optical loss and crosstalk noise on SP-NN power consumption (i.e., laser power penalty) as well as scalability constraints in SP-NNs. Finally, Section \ref{sec:: 6-Conclusion} concludes this work.
\section{Background and Prior Related Work}\label{background}

In this section, we present an overview of the MZI building block in coherent SP-NNs as the primary vector-matrix multiplier unit and some fundamentals of MZI-based coherent SP-NNs. We also discuss different sources of optical loss and crosstalk in MZIs. Moreover, we review prior work on studying the effect of loss and crosstalk in SP-NNs.

\subsection{Mach--Zehnder Interferometer (MZI)}
\begin{figure}[t]
\centering
\includegraphics[width=3.5in]{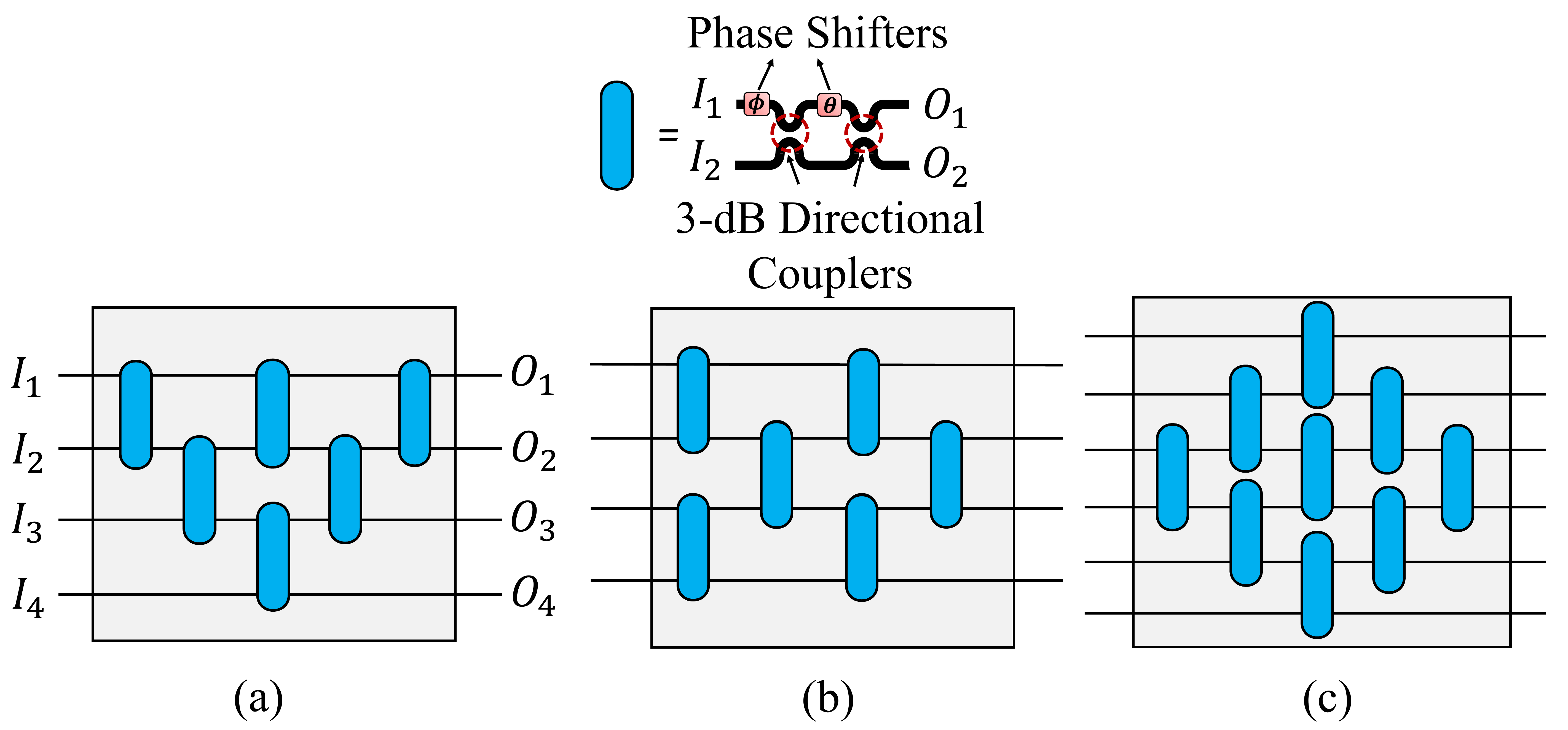}
  \caption{ Schematic of a 4$\times$4 single MZI-based optical interface unit (top) with different mesh configurations. (a) Reck, (b) Clements, and (c) Diamond.}\label{Fig:: Architectures}
  \vspace{-0.15in}
\end{figure}

MZIs can be used to realize linear multiplication between a 2$\times$1 vector (signals applied to the two inputs) and a 2$\times$2 matrix (defined based on the phase settings in the MZI). Such an MZI-based multiplier unit can be constructed using two 3-dB directional couplers (DCs) with an ideal splitting ratio of 50:50 and two integrated phase shifters ($\theta$ and $\phi$), as shown in Fig. \ref{Fig:: Architectures}. Phase shifters in this design can be implemented, for example, using microheaters on top of the underlying waveguide \cite{Farhad_4by4}. By introducing a temperature change using microheaters, the refractive index of the underlying silicon waveguide will change due to the thermo-optic effect, leading to a change in the phase of the electric field of the propagating optical signal. Therefore, by controlling the phase shift between the two arms in an MZI, we can control the interference in the output. Note that in the MZI in Fig. \ref{Fig:: Architectures}, $0\leq\theta\leq\pi$ and $0\leq\phi\leq2\pi$. The transfer matrix of an MZI-based multiplier unit can be realized by multiplying the transfer matrices of the two 3-dB DCs ($T_{DC1,DC2}$) and the transfer matrices of the two phase shifters ($T_{\phi,\theta}$). Accordingly, the ideal transfer matrix of an MZI-based multiplier unit (i.e., without optical loss and crosstalk noise) can be defined as\cite{shokraneh2020diamond,hamdani2023modelling}:
\begin{align*} 
    T_{MZI}(\theta,\phi) &= T_{DC2} \cdot T_{\theta} \cdot T_{DC1} \cdot T_{\phi} \\
    &= \begin{pmatrix}
      \frac{e^{i\phi}}{2}(e^{i\theta} - 1) & \frac{i}{2}(e^{i\theta} + 1)  \\
      \frac{ie^{i\phi}}{2}(e^{i\theta} + 1) & -\frac{1}{2}(e^{i\theta} - 1)
      \end{pmatrix}. \tag{1}
      \label{MZI}
\end{align*}

While MZIs can help perform matrix-vector multiplication in the optical domain, they have a large footprint. 
For example, state-of-the-art MZI multipliers can be up to about 300~$\mu$m long and 30$~\mu$m wide \cite{mirza2022characterization}, limiting the scalability of MZI-based coherent SP-NNs \cite{Farhad_4by4,shafiee_future}. 
In addition to a large footprint, they also suffer from high optical losses and crosstalk noise \cite{bahadori2016loss}. The optical loss in an MZI originates from the absorption in the metallic contacts in proximity when using microheaters for applying the required phase shifts on the two MZI arms, the directional couplers imperfections, and the propagation loss of the waveguides, which is mainly due to the sidewall roughness in the waveguides. The optical loss and crosstalk noise in MZIs will be discussed in detail in Section \ref{Loss_XT_framework}.

\subsection{MZI-based Coherent SP-NNs}\label{SP-NN-back}
MZI-based SP-NNs rely on the manipulation
of the electrical field's phase of a single optical wavelength to perform matrix-vector multiplication. MZIs in such SP-NNs are responsible for phase manipulations and interference to carry out the computations \cite{sunny2021survey, Clements:16, reck1994experimental,shokraneh2020diamond, Farhad_4by4}. MZIs can be cascaded in the form of an array following a specific configuration to implement an OIU for performing large matrix-vector multiplication in the optical domain. Fig. \ref{Fig:: MZI-Structure} shows an overview of a coherent SP-NN composed of an optical interface unit (OIU), optical gain unit (OGU), and nonlinear activation unit (NAU). A fully connected layer ($L_i$) with $n$ inputs performs matrix-vector multiplication between the input vector ($I_i$) and a weight matrix ($W_i$). The output vector of the OIU then will be passed into the OGU to realize unitary matrices with arbitrary magnitudes, and eventually into the NAU to apply a nonlinear activation function ($f_i$), the result of which will be the input to the next layer ($L_{i+1}$). The output of $L_i$ can be mathematically modeled as $O_{i}^{n_i \times 1} = f_i(W_i^{n_{i} \times n_{i-1}}, O_{i-1}^{n_{i-1} \times 1})$, in which $W_i$ is the layer's corresponding weight matrix. The weight matrix ($W_i$) can be obtained by training the network. Using singular value decomposition (SVD), the obtained weight matrix can be mathematically modeled as $W_i = U_i^{n_i \times n_i}\Sigma_i^{n_i \times n_{i-1}}V_i^{H, n_{i-1} \times n_{i-1}}$. In this formulation, $U_i$ and $V_i$ are unitary matrices. Moreover, $V_i^H$ denotes the Hermitian transpose of $V_i$ and $\Sigma_i$ is a diagonal matrix consisting of the eigenvalues of $W_i$. A unitary matrix can be implemented by an array of cascaded 2$\times$2 MZIs in a specific configuration according to:
\begin{equation}
U_i^{n_i \times n_i}=D\left(\prod_{(m, n) \in S} T_{MZI_{{m, n}}}\right). \tag{2} \label{SP-NN}
\end{equation}
Using this scheme, each unitary matrix $U$ can be decomposed into the products of several MZIs' transfer matrices. The order of the multiplication of MZIs' transfer matrices plays an important role in SP-NNs, determining the configuration of the MZIs in the SP-NNs (i.e. Clements, Reck, or Diamond see Fig.~\ref{Fig:: Architectures}). In \eqref{SP-NN}, $D$ is a diagonal matrix with complex elements with a unity modulus \cite{Clements:16}, and $S$ denotes the order of the multiplication of the MZIs' transfer matrices. $S$ will be determined based on the configuration of the array of cascaded MZIs used to map the weight matrices in order to perform matrix-vector multiplication in the optical domain.
Moreover, $m$ and $n$ denote the input ports (i.e., $I_1-I_4$ in Fig.\ref{Fig:: Architectures} (a)) which require transformation using each MZI in the network. For example, $m=1$ and $n=2$ refer to the MZI in between input ports 1 and 2 in the network. Note that $n=m+1$ always applies \cite{Clements:16}.  The location of each MZI in the network can be determined during the mapping of the weights to the array of cascaded MZIs which itself depends on the configuration of the network.
\begin{figure}[t]
\centering
\includegraphics[width=3.5in]{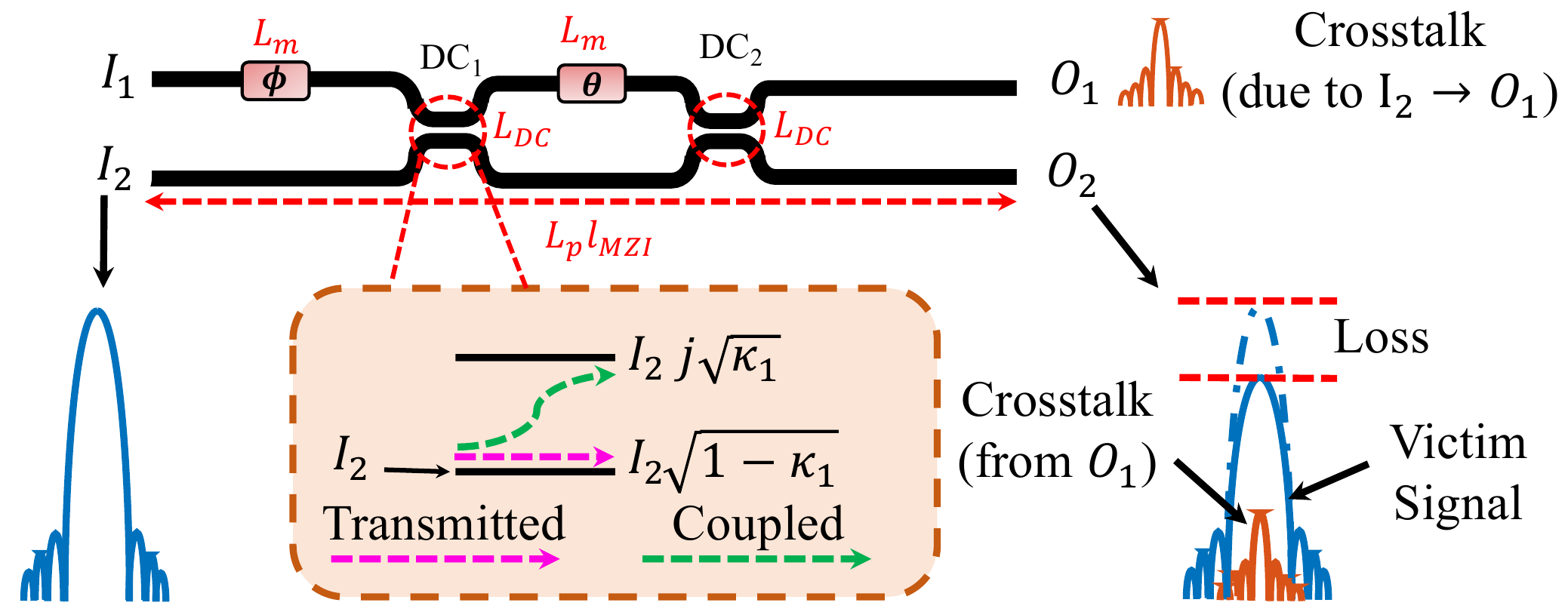}
  \caption{ Schematic of a 2$\times$2 MZI multiplier unit with different sources of optical loss and crosstalk noise. Here, $I_2\rightarrow O_2$ is shown as an example with $\theta=\pi$ ($l_{MZI}$: MZI length, $L_m$: loss due to metallic absorption, $L_{DC}$: DC's insertion loss, $L_p$: propagation loss). }\label{Fig:: MZI_loss_XT}
  \vspace{-0.15in}
  \label{Fig::MZI_loss_Xtalk}
\end{figure}

Several configurations have been proposed for the network topology of cascaded MZIs and to carry out the unitary transformation of the input optical signals. Three well-known mesh configurations: Reck, Clements, and Diamond \cite{reck1994experimental,Clements:16,shokraneh2020diamond} are considered in this paper to realize MZI-based unitary multipliers. A 4$\times$4 Reck mesh configuration is shown in Fig. \ref{Fig:: Architectures}(a) organizing an array of MZIs in a triangular shape. In general, any $N\times N$ unitary multiplier based on the Reck design consists of $\frac{N(N-1)}{2}$ MZIs, where $N$ is the number of the input ports. The same number of MZIs can be configured in a rectangular shape as is shown in Fig.~\ref{Fig:: Architectures}(b), and this configuration is called the Clements mesh. The advantage of the Clements design over the Reck is that the network is more symmetric, hence making the unitary multiplier more resilient to propagation loss due to more symmetric and, on average, shorter optical paths compared to the Reck design \cite{Clements:16}. We can also increase the number of MZIs to make the Reck design more symmetric. The work in \cite{shokraneh2020diamond} proposed this symmetric design by adding additional $\frac{(N-2)(N-1)}{2}$ MZIs to the Reck configuration to design the Diamond mesh configuration, shown in Fig. \ref{Fig:: Architectures}(c). For an array of cascaded MZIs with $N$ inputs, $(N-1)^2$ MZIs will be used in a diamond shape. Although the network topology includes a higher number of input and output ports, only the last $N$ inputs are used to perform matrix-vector multiplication, while the rest of the inputs can be used for characterization and calibration of the MZIs in the network \cite{shokraneh2020diamond}.

\subsection{Optical Loss and Crosstalk Noise in MZIs}
MZIs intrinsically suffer from optical loss and crosstalk noise. The schematic of a single 2$\times$2 MZI multiplier is shown in Fig. \ref{Fig:: MZI_loss_XT}. An optical signal traversing an MZI can undergo different losses based on the MZI phase settings. The main sources of optical loss in an MZI are the DCs, metal absorption---which varies slightly with the adjusted phase settings in the MZI---due to the proximity of microheater's metallic contacts to waveguides, and propagation loss mainly due to sidewall roughness in waveguides \cite{shang2020investigation}. In addition, crosstalk noise can originate from the undesired coupling of light in the DCs in an MZI. In coherent SP-NNs, we deal with coherent (i.e., intra-channel or in-band) crosstalk. As shown in Fig. \ref{Fig:: MZI_loss_XT}, coherent crosstalk noise can interfere with the main optical signal (victim signal) based on their phase difference, imposing power fluctuations on the victim signal. Unlike incoherent networks, in coherent networks like MZI-based SP-NNs, the in-band coherent crosstalk noise cannot be easily filtered in the output due to the coherence between the crosstalk and victim signals (i.e., on the same wavelength). 

Unlike in SP-NNs, optical loss and crosstalk noise have been widely studied in chip-scale Datacom photonic networks (e.g., \cite{Nikdast_crosstalk} and \cite{app10238688_Xtalk}), showing signal integrity degradation and scalability constraints in these networks due to optical loss and crosstalk noise. Unfortunately, the existing work on optical loss and crosstalk analysis in such networks cannot be applied to SP-NNs  because optical loss and crosstalk noise characteristics of silicon photonic devices for photonic computation in SP-NNs are different. For example, a 2$\times$2 MZI switching cell, whose structure is similar to the one in Fig.~\ref{Fig:: MZI_loss_XT} but without $\phi$, in an optical switch fabric can only take two functional states based on $\theta$ for optical loss and crosstalk analysis: the Cross-state, where $\theta=0$ and $I_1\rightarrow O_2$ and $I_2\rightarrow O_1$, and the Bar-state, where $\theta=\pi$ and $I_1\rightarrow O_1$ and $I_2\rightarrow O_2$. As a result, crosstalk noise can be easily characterized in such a device. However, in the coherent SP-NNs, $\theta$, which determines the MZI state, can take any value between 0 and $\pi$ (0$\leq\theta\leq\pi$) to perform computation via interference between the inputs. The analysis of optical loss and crosstalk in SP-NNs should therefore account for various phase settings in the underlying MZI devices. 

\subsection{Prior Related Work}
While several coherent SP-NNs have been recently proposed \cite{PourFard:20,shokraneh2020diamond}, the work in \cite{banerjee2021modeling, mirza2022characterization} showed that the inferencing accuracy of such networks can drop by up to 90\% due to fabrication-process variations and thermal crosstalk. In addition to such variations, the work in \cite{shokraneh2020diamond} explored the impact of optical loss and phase noise in MZIs for the Diamond SP-NN configuration and showed that the SP-NN's inferencing accuracy can drop to below 20\% when scaling the network in the presence of phase errors and MZI losses. However, the work in \cite{shokraneh2020diamond} simply assumed that the loss in different devices in a network to be normally distributed (irrespective of their phase settings).

The work in \cite{liu2022reliability} systematically analyzed the impact of loss and crosstalk in SP-NNs. However, the loss and crosstalk models presented in \cite{liu2022reliability} are not a function of the phase settings of the MZIs. Hence, only the maximum and minimum loss and crosstalk are calculated. In addition, the models illustrated in \cite{liu2022reliability} cannot be extended to OIUs of any arbitrary configuration.
As the size and complexity of emerging SP-NNs increase to handle more complex tasks, the total insertion loss accumulated in the network increases as well. This necessitates the use of power-hungry optical amplification devices \cite{haq2020micro_SOA_Cband} and higher laser power at the input. Uncertainties due to fabrication-process variations---the analysis of which is beyond the scope of this paper---in the two DCs in an MZI can degrade the extinction ratio (ER) of the device which, in turn, will increase the loss and crosstalk in the output \cite{De_marinis_app11136232}. Yet, no prior work comprehensively analyzes the impact of optical loss and crosstalk noise in coherent SP-NNs from the device to the system level. Although the work in \cite{De_marinis_app11136232} suggests that using silicon nitride instead of silicon to implement MZI-based SP-NNs as a possible solution to reduce losses, the performance degradation due to coherent crosstalk in SP-NNs still remains unaddressed.\par

Different from the aforementioned work, this paper presents a comprehensive modeling framework for the optical loss and coherent crosstalk noise in coherent SP-NNs. The proposed loss and crosstalk analysis at the MZI device level takes into account the phase settings on the device. In addition, the models developed at the network level are adaptable meaning that the number of inputs/outputs and layers in SP-NNs can vary to explore SP-NN optical power penalty and scalability constraints while evaluating average and worst-case optical loss and crosstalk. Compared to our prior work in \cite{shafiee2022loci}, we extended our analysis for two more well-known SP-NN configurations (Reck and Diamond) in addition to the Clements configurations. We show that our proposed models can be applied to any SP-NN with any configuration, proving their versatility. We present a comprehensive analysis of the laser power penalty and inferencing accuracy loss in SP-NNs due to optical loss and crosstalk noise. In addition, we present an analysis of the effect of optical loss and crosstalk when optoelectronic NAUs are used in the SP-NNs. We also present a detailed analysis of scalability constraints of SP-NNs due to optical loss and crosstalk when a single OIU is being used as a processing unit to carry out the computations in the optical domain.

\section{Optical Loss and Crosstalk Noise Analysis}\label{Loss_XT_framework}
This section presents the compact models developed to analyze optical loss and crosstalk noise in MZI-based coherent SP-NNs from the device level to the network level. 
We also model the loss and crosstalk noise in the optoelectronic nonlinear activation unit in coherent SP-NNs. All these models will be used to explore the power penalty, performance (e.g., inferencing accuracy), and scalability constraints in SP-NNs of different sizes and architectures under optical loss and crosstalk noise, as will be discussed in Section \ref{Simulation}.

\subsection{Modeling Optical Loss and Crosstalk at Device Level}
The schematic of a single 2$\times$2 MZI-based multiplier unit is shown in Fig. \ref{Fig:: MZI_loss_XT}. It comprises two 3-dB DCs and two integrated phase shifters ($\theta$ and $\phi$) on the upper arm. DCs in the MZI structure are responsible for splitting and combining the optical signals traversing the MZI. Optical crosstalk noise stems from undesired mode coupling in these DCs in the MZI structure.  The splitting ratio for an optical signal at the input of a DC can be determined by the cross-over coupling coefficient ($\kappa$) and the power transmission coefficient ($t$) in the DC. Considering the DC optical loss to be $L_{DC}$, we have $|\kappa|^2 +|t|^2=L_{DC}^2$. To be used in an MZI-based multiplier unit for coherent SP-NNs, the DCs should perform exact 50:50 splitting/combining ($\kappa = t =$~0.5). Both
$\kappa$ and $t$ are wavelength dependent and they also depend on the waveguide width and thickness and the gap between the waveguides in DCs. As discussed in Section \ref{background}, the main sources of optical loss in an MZI are the DC's loss ($L_{DC}$), propagation loss ($L_p$), and the metal absorption loss ($L_m$) due to interaction of the propagating light with the metallic contacts integrated on top of the waveguide when using microheaters to implement the phase shifters. The propagation loss originates from sidewall roughness and scattering loss in the waveguides \cite{Bahadori:16}. The metal absorption loss depends on the metal used to implement the heater's contacts and their geometry such as thickness, width, length, and the longitudinal distance between the waveguide and the contacts \cite{ding2016broadband}.

Considering \eqref{MZI} and the MZI's optical loss parameters defined (i.e., $L_{DC}$, $L_m$, and $L_p$, as shown in Fig. \ref{Fig:: MZI_loss_XT}), an optical-loss-aware transfer matrix model for the 2$\times$2 MZI multiplier unit can be defined as: 
\begin{align*}
\begin{pmatrix}
O_{1} \\
O_{2}
\end{pmatrix}&= 
 T_{DC_{2}} \cdot T_{\theta} \cdot T_{DC_{1}} \cdot T_{\phi} \cdot \begin{pmatrix}
I_{1} \\
I_{2}
\end{pmatrix}\label{eq::MZI_trans_mat}
\end{align*}
\begin{align*}
T_{DC_{2}}&=\begin{pmatrix}
L_{DC}\sqrt{1-\kappa_{2}} & L_{DC}j\sqrt{\kappa_{2}} \\
L_{DC}j\sqrt{\kappa_{2}} & L_{DC}\sqrt{1-\kappa_{2}}
\end{pmatrix},
\end{align*}
\begin{align*}
T_{\theta}=\begin{pmatrix}
L_{p}l_{MZI}L_{m}e^{j\theta} & 0 \\
0 & L_{p}l_{MZI}
\end{pmatrix},
\end{align*}
\begin{align*}
T_{DC_{1}}&=\begin{pmatrix}
L_{DC}\sqrt{1-\kappa_{1}} & L_{DC}j\sqrt{\kappa_{1}} \\
L_{DC}j\sqrt{\kappa_{1}} & L_{DC}\sqrt{1-\kappa_{1}}
\end{pmatrix},
\end{align*}
\begin{align*}
T_{\phi}=\begin{pmatrix}
L_{m}e^{j\phi} & 0 \\
0 & 1
\end{pmatrix}.\tag{3}
\end{align*}
In this model, $\kappa_1$ and $\kappa_2$ are the cross-over coupling coefficients in the two DCs. 

Optical crosstalk noise in a 2$\times$2 MZI switching element, whose structure is similar to the one shown in Fig.~\ref{Fig:: MZI_loss_XT} but without $\phi$ phase shifter at the input of the MZI, can be analyzed by considering the MZI in the Bar ($\theta=\pi$) or in the Cross-state ($\theta=$~0). By injecting an optical signal into one of the input ports of the MZI, the crosstalk noise can be determined by capturing the undesired optical signal transmission on the opposite output when in Cross- or Bar-states. For example, considering the MZI schematic depicted in Fig. \ref{Fig:: MZI_loss_XT}, when injecting light into $I_2$, by setting $\theta=\pi$, the crosstalk coefficient in the Bar-state ($X_B$) can be captured on $O_1$. Similarly, it is possible to capture the Cross-state crosstalk coefficient ($X_C$) on $O_2$ by setting $\theta=$~0. Different from the MZIs used in switching networks, MZIs in coherent SP-NNs can take more than the two Bar-state and Cross-state, depending on the value of $\theta$. In a 2$\times$2 MZI-based multiplier unit, $\theta$ can assume any value between 0 and $\pi$ (0$\leq\theta\leq\pi$) to perform a 2$\times$2 unitary multiplication ($0\leq\phi\leq2\pi$ does not change the MZI state). Therefore, the exact analysis of the crosstalk noise per device's output ports cannot be easily performed in coherent SP-NNs.

To address this limitation, we can define a statistical model to estimate the crosstalk noise on the output ports of each MZI in coherent SP-NNs depending on the MZI's $\theta$ phase setting. Considering the crosstalk coefficient in the two known Cross-state and Bar- state ($X_B$ and $X_C$, where typically $X_B\leq X_C$ \cite{Shoji:10Xtalk}), we can statistically model the crosstalk coefficient in the MZI in any intermediate state ($X$) as the function of $\theta$ using a Gaussian distribution whose mean can be calculated according to $\mu(\theta)=\frac{X_B - X_C}{\pi}\theta+X_C$ and standard deviation of 0.05$\cdot\mu(\theta)$, considered here as an example. 
Using the crosstalk coefficient in intermediate states $X$, the transfer matrix of a 2$\times$2 MZI-based multiplier unit in \eqref{eq::MZI_trans_mat} can be written as:
\begin{align*}
\begin{pmatrix}
O_{1} \\
O_{2}
\end{pmatrix}=\begin{pmatrix}
(1-X)T_{11} & (1-X)T_{12} \\
(1-X)T_{21} & (1-X)T_{22}
\end{pmatrix}
\begin{pmatrix}
I_{1} \\
I_{2}
\end{pmatrix}\\
+\begin{pmatrix}
(X)T_{21} & (X)T_{22} \\
(X)T_{11} & (X)T_{12}
\end{pmatrix}
\begin{pmatrix}
I_{1} \\
I_{2}
\end{pmatrix}\tag{4}\label{eq::MZI_xtalk}.
\end{align*}
\begin{figure}[t]
\centering
\includegraphics[width=3.5in]{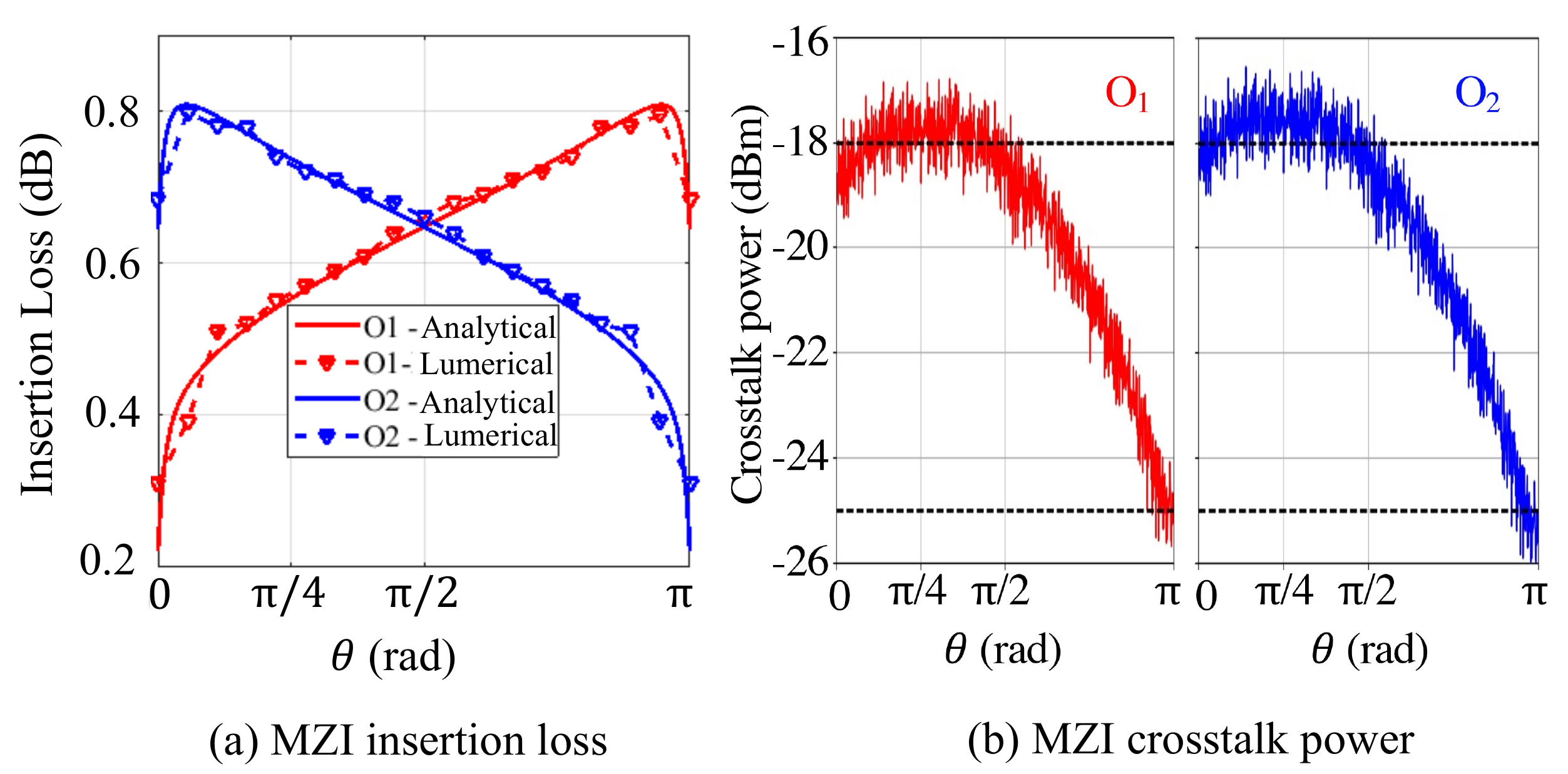}
  \caption{ Insertion loss, (a), and crosstalk power, (b), at the output of the 2$\times$2 MZI in Fig.~\ref{eq::MZI_xtalk} simulated using the parameters listed in Table~\ref{Table:: Table1}.}\label{fig::device_results}
\end{figure}

Models presented in \eqref{eq::MZI_trans_mat} and \eqref{eq::MZI_xtalk} can be used to analyze the optical loss and crosstalk noise in any MZI with different phase settings. For example, considering the parameters listed in Table \ref{Table:: Table1}, Fig.~\ref{fig::device_results} shows the insertion loss and crosstalk power for a single 2$\times$2 MZI multiplier unit.
In this figure, the x-axis shows the  state of the MZI based on $\theta$ where $0\leq\theta\leq\pi$. Note that $\phi$ does not change the MZI state, but its loss is included in the results shown.  Observe that the insertion loss on each output port is $\approx$0.3--0.8~dB.
We used Ansys Lumerical \cite{Lumerical} to validate the results in Fig.~\ref{fig::device_results}(a). Note that commercial tools (e.g., Lumerical or Synopsys) cannot analyze the crosstalk in intermediate states in the MZI, hence  their results are not considered in Fig.~\ref{fig::device_results}(b). 
 Let us revisit Fig.~\ref{Fig:: MZI_loss_XT}: compared to input $I_2$, the optical signal on $I_1$ experiences higher insertion loss because of $L_{m}$ through $\phi$. Therefore, for example, the insertion loss is higher on $O_2$ for the Cross-state when $\theta=$~0, and it is higher on $O_1$ for the Bar-state when $\theta=\pi$. Note that the fluctuations in the crosstalk power in Fig. \ref{fig::device_results} are due to the Gaussian noise model defined for the MZI. The coherent crosstalk power in the MZI output changes between $\approx-$18~dBm and $\approx-$25~dBm, when the input power is 0~dBm.

\subsection{Modeling Optical Loss and Crosstalk at Network Level}
As discussed in Section \ref{SP-NN-back}, the weight matrix of each hidden layer $M$ in an SP-NN architecture with $N_1$ inputs and $N_2$ outputs can be decomposed into a multiplication of two unitary and one diagonal matrix implemented by OIUs followed by an OGU, which can be implemented using semiconductor optical amplifiers (SOAs), and an NAU, as shown in Fig.~\ref{Fig:: MZI-Structure}. The number of MZIs in an OIU depends on the mesh configuration (Clements, Reck, or Diamond) and the number of inputs ($N_1$) and outputs ($N_2$). Moreover, as discussed in Section III-A, the optical loss and crosstalk noise at the outputs of an MZI in an OIU is a function of $\theta$ and $\phi$ in the MZI, which can be obtained during the training of the network \cite{shafiee2022loci}.

The optical loss in the output of an SP-NN layer ($L_i$) can be systematically modeled as:

\begin{equation*}
IL_{i}=  G \cdot IL_{OIU}(C, N_1, N_2, \{\theta\}, \{\phi\}) \cdot IL_{NAU}.\tag{4}\label{eq::layer_loss}  
\end{equation*}
In \eqref{eq::layer_loss}, $IL_{OIU}$ is the insertion loss in the OIU that can be calculated using \eqref{eq::MZI_trans_mat} for each MZI, and it depends on the OIU's mesh configuration ($C$), its dimension ($N_1$ and $N_2$), and the phase settings of MZIs in the OIU. Moreover, $G$ is the optical gain of the SOAs in the OGU, and $IL_{NAU}$ is the insertion loss due to the NAU. In this paper, we consider the state-of-the-art SOA in \cite{haq2020micro_SOA_Cband} with $G=$~17~dB. Also, the insertion of the optoelectronic NAU is considered to be $IL_{NAU}=$~0--1~dB, depending on its nonlinear response \cite{PourFard:20}.

Optical crosstalk noise from each MZI will be accumulated at the outputs of the OIU as the optical signal propagates through the SP-NN. Therefore, following the same approach for optical loss calculation, the accumulated optical crosstalk noise power at the end of the $L_i$ can be modeled as:
\begin{equation*}
 XP_i = \sum_{j=1}^{N_{MZI}} \left(P \cdot X^{ij}_{MZI}(\rho)\cdot IL^{ij}_{OIU} \right) \cdot G \cdot
IL_{NAU}.\tag{5}\label{eq::layer_xtalk}\vspace{-0.05in}    
\end{equation*}
In \eqref{eq::layer_xtalk}, $N_{MZI}$ is the total number of MZIs in the OIU in layer $M_i$ and it depends on the configuration of the OIU as well as its dimension. Also, $P$ is the input optical power. Moreover, $X^{ij}_{MZI}(\rho)$ can be calculated using \eqref{eq::MZI_xtalk} and is the coherent crosstalk coefficient on the output of layer $L_i$ originating from MZI$_j$ in the OIU. Also, $\rho$ is the optical phase of the crosstalk signal. Similarly, $IL^{ij}_{OIU}$ is the power attenuation fraction coefficient per MZI, which can be calculated using \eqref{eq::MZI_trans_mat}, experienced by $X^{ij}_{MZI}(\rho)$ as it traverses the OIU. While SOA's gain can compensate for the OIU insertion loss (to some extent), SOAs can also amplify the optical coherent crosstalk noise on the SP-NN's outputs. By integrating the insertion loss and crosstalk models in \eqref{eq::layer_loss} and \eqref{eq::layer_xtalk} across multiple layers, we can analyze the network-level optical insertion loss and crosstalk power in coherent SP-NNs of any size and configuration.

\subsection{Modeling Optical Loss and Crosstalk in Optoelectronic NAUs}\label{NAU_loss}
Nonlinear activation functions are an integral part of deep neural networks due to the essential need for the realization of the complex nonlinear relationship between the inputs and outputs of the SP-NNs \cite{PourFard:20}. NAUs are responsible to trigger a single activation at the end of each layer's output and pass the output to the input of the next layer. 
NAUs can be implemented electronically \cite{SiPh_codesign}, optoelectronically \cite{PourFard:20}, or optically \cite{shen2017deep_nature}, each with different costs. High power consumption and latency as well as the need for lasers because of the need for multiple E-O and O-E conversions can be named as limitations related to electronically implemented NAUs. Moreover, very large waveguide lengths and high optical power must be used for optical NAUs due to the weak nonlinearity of photonic platforms \cite{PourFard:20, destras2023survey}. 
Optoelectronic NAUs presented in \cite{PourFard:20,williamson2019reprogrammable} show great promise as alternatives to electronic ones due to the ability to implement an arbitrary nonlinear response via self-intensity modulation (e.g., MZI-based electro-optical modulators) of the input optical signal. Note that research on optical NAUs is still ongoing. Therefore, in most cases, optoelectronic or electronic NAU are used in SP-NNs.
 
\begin{figure}[t]
\centering
\includegraphics[width=3.5in]{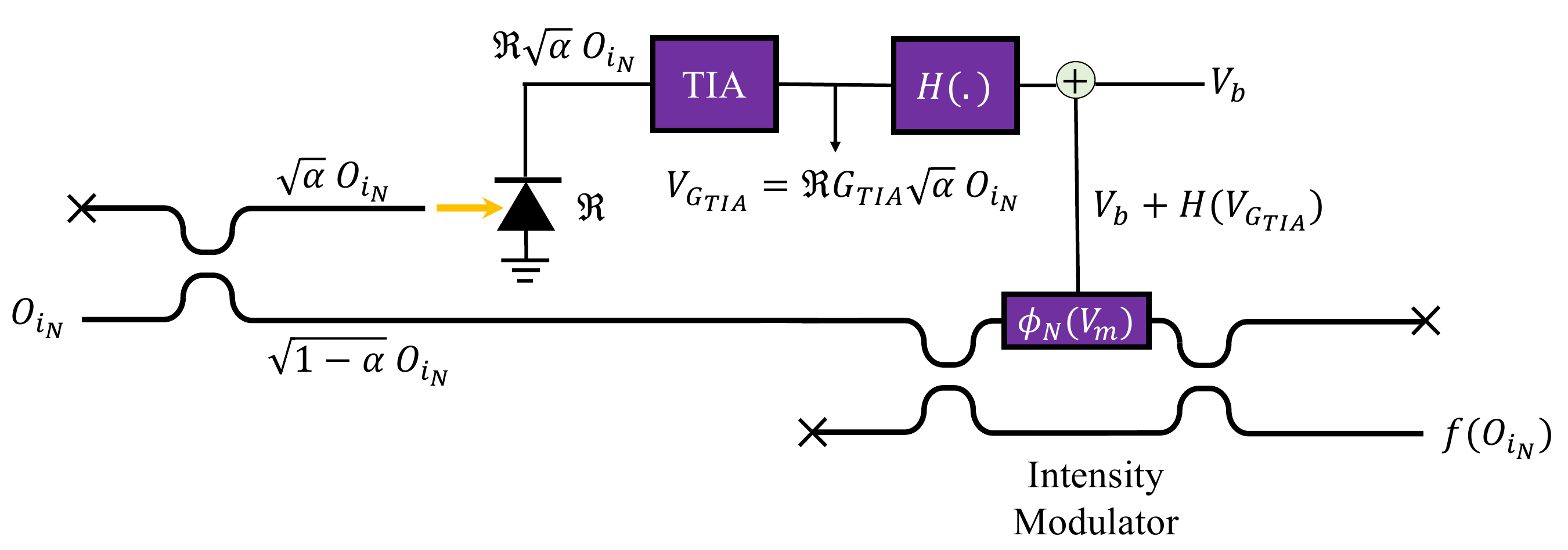}
  \caption{Schematic block diagram of the optoelectronic nonlinear activation unit used to realize ReLU-like nonlinear activation function \cite{PourFard:20, williamson2019reprogrammable}. }\label{Fig:: NAU}
  \vspace{-0.15in}
\end{figure}

 The schematic diagram of the optoelectronic NAU considered in our work to realize \textit{ReLU}-like activation response is shown in Fig.~\ref{Fig:: NAU}. The optical signal at each output of the OIU ($O_{i_{N}}$) will pass through a directional coupler with a cross-over coupling coefficient of $\alpha$ ($\alpha=$~0.1 in this paper, see Table \ref{table_1}). The larger portion of the split power ($\sqrt{1-\alpha} O_{i_{N}}$) will go through an intensity modulation using an MZI-based electro-optical modulator (see Fig. \ref{Fig:: NAU}). The remaining portion of the input optical power ($\sqrt{\alpha} O_{i_{N}}$) is used to drive the MZI-based optical modulator used in the NAU \cite{PourFard:20}, based on the following principle: the optical signal $\sqrt{\alpha} O_{i_{N}}$ routed in the upper branch enters a photodetector (PD) with the responsivity of $\Re$~(A/W)---$\Re \times P_{in} = I_{PD}$ for an ideal PD where $I_{PD}$ is the PD's photocurrent---followed by a transimpedance amplifier (TIA) with a gain of $G_{TIA}$, to apply a gain to the output of the PD and convert the PD's photocurrent to a voltage output. The output of the TIA will then goes through a signal conditioning unit ($H(.)$) to apply an arbitrary function to the output voltage of the TIA. In our analysis, we used the identity function $H(\Re G_{TIA}\sqrt{\alpha} O_{i_{N}}) = \Re G_{TIA}\sqrt{\alpha} O_{i_{N}}$ for the sake of simplicity as the conditioning unit. 
 Finally, an additional bias voltage ($V_b$) and the output of the conditioning unit will be used to drive the MZI-based electro-optical intensity modulator in the lower branch. 
 
 The ideal nonlienarity ($f(O_{i_{N}})$) of the MZI-based optoelectronic NAU depicted in Fig. \ref{Fig:: NAU} can be mathematically modeled as \cite{williamson2019reprogrammable, PourFard:20}:
\begin{align*}
 f(O_{i_{N}})=  & 
  j O_{i_{N}}\sqrt{1-\alpha} \cos \left(\pi \frac{V_b}{2V_{\pi}} + \pi \frac{H\left(G_{TIA}\Re\alpha|O_{i_{N}}|^2\right)}{2V_{\pi}}\right)\\
 & \cdot \exp \left(-j\left[\pi \frac{V_b}{2V_{\pi}} + \pi \frac{H\left(G_{TIA}\Re\alpha|O_{i_{N}}|^2\right)}{2V_{\pi}}\right]\right).\tag{6} \label{Eq:: NAU_f}
\end{align*}
Here, $V_\pi$ is the voltage that is required to impose a $\pi$ phase shift in the MZI-based intensity modulator in the optoelectronic NAU. A \textit{ReLU}-like nonlinear activation response can be realized by setting $V_b=V_{\pi}$ in the formulation proposed in \eqref{Eq:: NAU_f}. Note that in the activation function modeled in \eqref{Eq:: NAU_f} and \cite{williamson2019reprogrammable, PourFard:20}, the effect of optical loss and crosstalk noise from the OIU unit, PD's sensitivity, shot noise, dark current (output current of the PD in the absence of any optical signal), and optical insertion loss of the DC and MZI in the NAU architecture are not considered. 

To update \eqref{Eq:: NAU_f} for analyzing the effect of optical loss and crosstalk noise in SP-NNs when optoelectronic NAUs are used, 
we can replace $O_{i_{N}}$ in \eqref{Eq:: NAU_f} with $IL_{OIU}\cdot G \cdot (O_{i_{N}} + XP_i(\theta_{err}))$. Note that in these formulations,  $G$ and the loss values are considered in terms of power amplification and attenuation coefficient. Moreover, $\theta_{err}$ is taking into account the phase of the crosstalk noise at the output of the OIU unit. Using this approach, we can write the input of the optoelectronic NAU unit including the optical loss and crosstalk from OIU ($O'_{i_{N}}$) as \cite{ghione2009semiconductor}:
\begin{equation*}   
O'_{i_{N}}=IL_{OIU}\cdot G\cdot (O_{i_{N}}+XP_i \exp (-j \theta_{err})). \tag{7}
\end{equation*}
The output photocurrent of the PD with a responsivity of $\Re$ and shot noise of $I_{shot}$ and dark current of $I_{dark}$ can be modeled as:
\begin{equation*}
I_{P D}=\Re O_{i_{N}}^{'2}+I_{s h o t}+I_{d a r k}, \tag{8}
\end{equation*}
where $I_{shot} = \sqrt{2 q I_{in}^2 \Re B}$ in which $B$ is the PD's bandwidth and $q=1.60\times 10^{-19}$ C is the electronic charge \cite{ghione2009semiconductor}. The output photocurrent then enters the TIA with the gain of $G_{TIA}$ and will be converted to an amplified voltage $V_G$ according to:

 \begin{equation*}
     V_{G_{TIA}} = G_{TIA}\cdot I_{PD}. \tag{9}
 \end{equation*}

Keeping the assumption of $V_H=H(V_{G_{TIA}})=V_{G_{TIA}}$, we can systematically model the activation function response using an MZI-based optoelectronic intensity modulator mentioned in \cite{williamson2019reprogrammable,PourFard:20} under optical loss and crosstalk noise from the OIU unit 
as:
\begin{align*}
  f_{n}(O_{i_{N}}) = &
 j L_{mod} O'_{i_{N}} \sqrt{1-\alpha}   \exp \left(-\frac{j \Delta \phi_N}{2}\right) \cos \left(\frac{\Delta \phi_N}{2}\right)\\
  & + L_{mod} \sqrt{1-\alpha} O'_{i_{N}} X_{mod}\cdot e^{j \theta_{mod}}\tag{10}\label{eq::NAU_loss_xt}
\end{align*}
where 
\begin{equation*}
\Delta \phi_N=\frac{\pi}{V_\pi}\left[V_b+V_H\right]. \tag{11}
\end{equation*}
 Here, we also assumed that $V_b=V_H$ to realize \textit{ReLU}-like activation response and the output photocurrent of the PD is considered to be zero ($I_{PD}=0$) for $O_{i_{N}}^{'2}\leq S_{PD}$, where $S_{PD}$ stands for PD sensitivity (the minimum detectable optical power by the PD). In \eqref{eq::NAU_loss_xt}, $X_{mod}$ is the optical crosstalk coefficient related to the MZI-based electro-optical modulator, $L_{mod}$ is its insertion loss, and $\theta_{mod}$ is 
  the phase of the optical crosstalk noise from the MZI in the intensity modulator in the optoelectronic NAU \cite{williamson2019reprogrammable}.

\begin{table}[t]
    \centering
    \caption{Device-level loss, crosstalk coefficient, power, gain, and NAU parameters considered in this paper (PhS: Phase shifter, PD: Photodetector).}
    \label{table_1}

\begin{tabular}{|c|c|c|c|} 
 \hline
 \rowcolor{Gray}
 Par. & Definition & Value & Ref. \\ [0.5ex] 
 \hline\hline
 $X_B$ & Crosstalk in Bar-state &  -25 dB & \cite{Shoji:10Xtalk}\\ 
 \hline
$X_C$ & Crosstalk in Cross-state & -18 dB & \cite{Shoji:10Xtalk}\\
 \hline
 $l_{MZI}$ & MZI length  & 300 $\mu$m & \cite{Farhad_4by4}  \\
\hline
$L_{m}$ & PhS (metal) absorption loss & 0.2 dB & \cite{ding2016broadband}\\
\hline
$L_{p}$ & Propagation loss & 2 dB/cm & \cite{Bahadori:16}\\
 \hline
 $L_{DC}$ & Insertion loss of DC & 0.1 dB & \cite{Bahadori:16}\\
 \hline
 $G$ & SOA gain & ~17 dB & \cite{haq2020micro_SOA_Cband}\\
 \hline
 $P$ & Input optical power & 0 dBm & -\\
 \hline
$G_{TIA}$ & TIA gain & $100$ $\Omega$ & \cite{PourFard:20}\\
 \hline
$\Re$ & Responsivity & 1 A/W & \cite{PourFard:20,fard2016responsivity,williamson2019reprogrammable}\\
 \hline
$\alpha$ & DC's splitting ratio in NAU& 0.1 & \cite{PourFard:20,fard2016responsivity,williamson2019reprogrammable}\\
 \hline
$V_{\pi}$ & Intensity modulator $\pi$ voltage & 10 V & \cite{PourFard:20,fard2016responsivity,williamson2019reprogrammable}\\
 \hline

$V_{b}$ & NAU's bias voltage & 10 V & \cite{PourFard:20,fard2016responsivity,williamson2019reprogrammable}\\
 \hline

$B$ & PD's bandwidth & 42.5 GHz & \cite{fard2016responsivity}\\
 \hline
$I_{dark}$ & PD's dark current & 3.5 $\mu$A & \cite{fard2016responsivity}\\
 \hline

 $S_{PD}$ & PD's sensitivity & -11.7 dBm & \cite{fard2016responsivity,9199100PD}\\
 \hline
\end{tabular} \label{Table:: Table1}
\end{table}

\section{Simulation Results and Discussions} \label{Simulation}


Optical loss and crosstalk noise lead to the deterioration of the performance of SP-NNs. We developed a framework to analyze the effect of optical loss and crosstalk noise in MZI-based coherent SP-NNs on top of {\fontfamily{qcr}\selectfont
Neuroptica
}\cite{shokraneh2020diamond, Neuroptica_gitlab}. {\fontfamily{qcr}\selectfont
Neuroptica
}is a flexible chip-level simulation platform for nanophotonic neural networks written in Python/NumPy. It provides a wide range of abstraction levels for simulating optical neural networks \cite{Neuroptica_gitlab}. We expanded the analysis of loss and crosstalk noise for a single MZI 
using the defined mathematical models in Section \ref{Loss_XT_framework} and parameters listed in Table \ref{table_1} to perform layer-level (i.e., OIU), network-level (i.e., multi-layer SP-NN), and system-level (i.e., network accuracy) analyses using {\fontfamily{qcr}\selectfont
Neuroptica
}\cite{Neuroptica_gitlab}. 
For layer- and network-level analyses, we consider random phase settings for MZIs in OIUs of different dimensions and configurations ($N=$~8, 16, 32, and 64, $M=$~1, 2, and
3 layers ). SVD is used to obtain the corresponding weight matrix in an SP-NN (see Fig. \ref{Fig:: MZI-Structure}) with three mesh configurations of Reck, Diamond, and Clements (see Fig. \ref{Fig:: Architectures}). Note that the random phases are only used for layer- and network-level analyses. As for the system-level analysis, we use shifted fast Fourier transform (shifted-FFT) on the MNIST handwritten digit dataset. The reason for using shifter-FFT is to reduce the number of inputs which leads to the size of the mesh configurations being smaller and more manageable so we can carry out the training of the SP-NNs of different configurations (i.e., Reck, Diamond, and Clements). Note that the training of the network is performed on complex inputs which leads to complex values of the weights when the training of the SP-NN is finished \cite{banerjee2022modeling}.
\begin{figure*}[t]
\centering
\includegraphics[width=7 in]{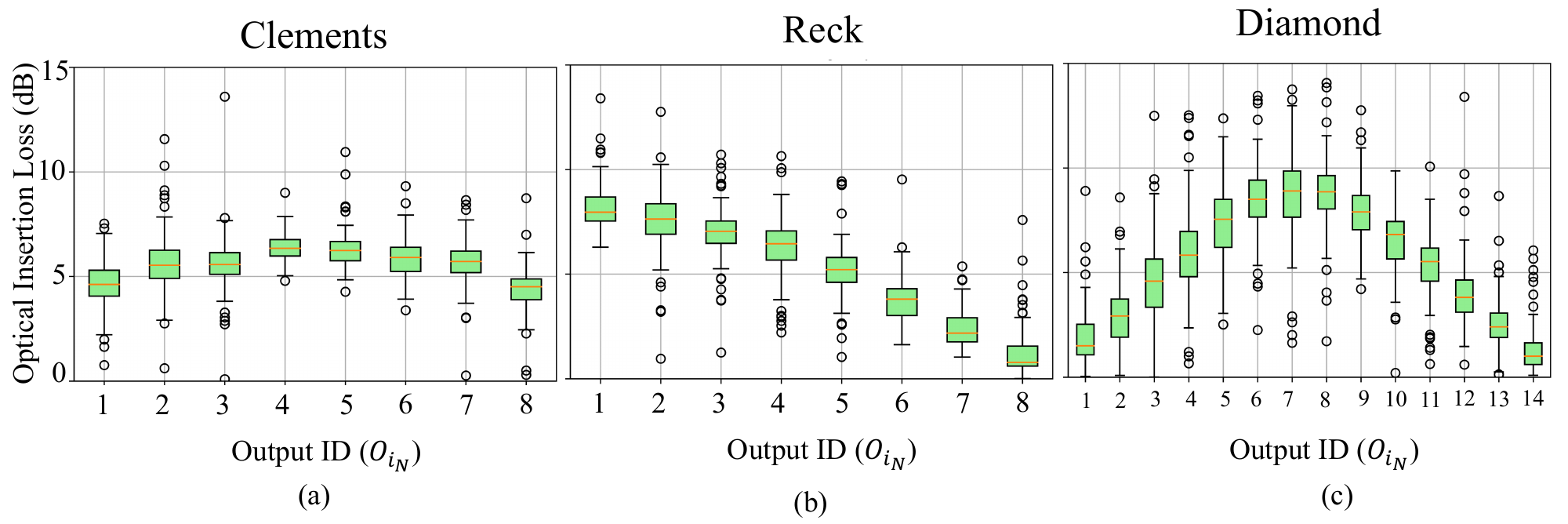}
  \caption{Optical loss analysis for 100 random weight matrices mapped to a fully connected 8$\times$8 OIU with different mesh configurations: (a) Clements, (b) Reck, and (c) Diamond. Output IDs are numbered from top to bottom (see Fig. \ref{Fig:: Architectures}).}\label{Fig:: Single_Layer_loss}
  \vspace{-0.15in}
\end{figure*}

\begin{figure*}[t]
\centering
\includegraphics[width=7.1 in]{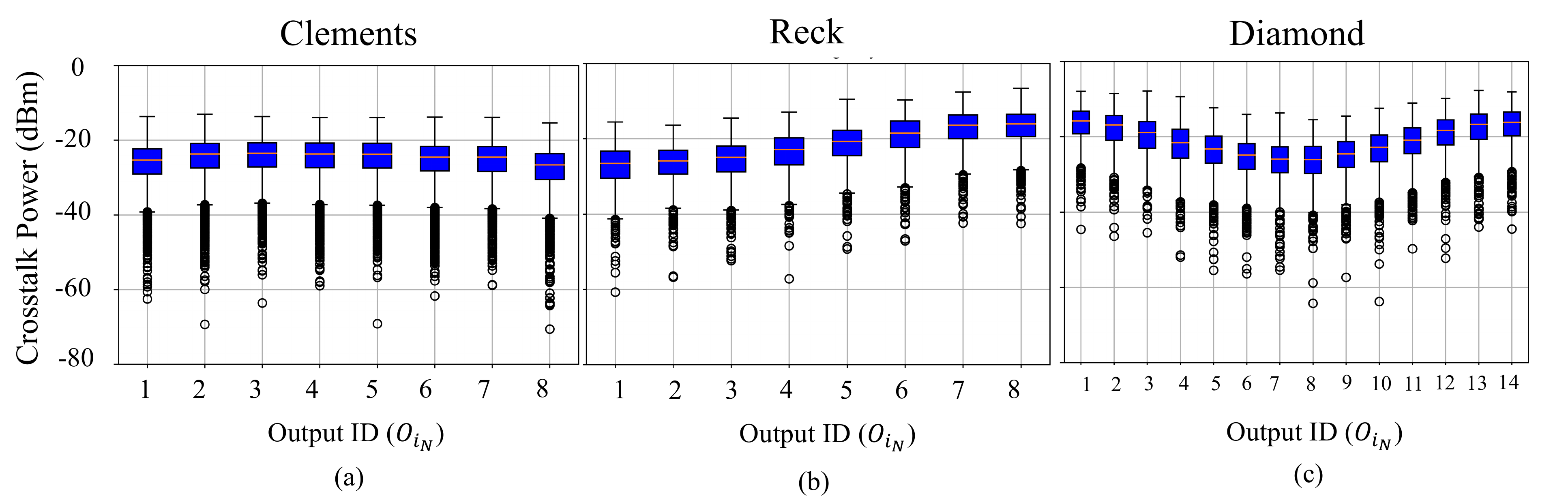}
  \caption{Optical crosstalk power analysis for 1000 random weight matrices mapped to a fully connected 8$\times$8 OIU with different mesh configurations: (a) Clements, (b) Reck, and (c) Diamond.  Output IDs are numbered from top to bottom (see Fig. \ref{Fig:: Architectures}). }\label{Fig:: Single_Layer_xt}
  \vspace{-0.15in}
\end{figure*}
We use inferencing accuracy as a figure of merit to analyze the effect of optical loss and crosstalk noise at the system level in the SP-NNs using the ideal \textit{ReLU} activation function. Moreover, we use relative-variation distance (RVD) (a measure of the deviation of two matrices \cite{banerjee2021modeling}) for the scalability analysis of OIUs of different scales with different mesh configurations. Such an analysis is helpful when using standalone OIUs as a photonic multiplication unit to perform matrix-vector multiplication. We also present a comprehensive analysis of the effect of optical loss and crosstalk noise on the SP-NNs' performance when optoelectronic NAUs are used instead of the ideal \textit{ReLU} activation response. In addition,  we analyze the laser power penalty in SP-NNs with different scales and configurations to compensate for optical loss and crosstalk noise.

\subsection{Optical Loss and Crosstalk in OIUs}
Using the analytical models in
\eqref{eq::layer_xtalk} and \eqref{eq::layer_loss} proposed in Section \ref{Loss_XT_framework}, a single layer ($M=1$) OIU with $N=$~8 (considered as an example), three different mesh configurations of Clements, Reck, and Diamond are simulated to capture the impact of optical loss and crosstalk noise at the OIU outputs. In these simulations, the optical insertion loss is calculated for 100 random weight matrices and the results are shown in the form of box plots in Fig.~\ref{Fig:: Single_Layer_loss}. We can see from Fig.~\ref{Fig:: Single_Layer_loss}(a) that for the Clements configuration, the average insertion loss among all the output ports is 6.5~dB, while the worst-case insertion loss can be as high as 14.8~dB. The average insertion loss for Reck and Diamond configurations is 6~dB and 6.5~dB, respectively. As for the worst-case loss, the Reck configuration experiences 15.11~dB, while the Diamond configuration undergoes 18~dB of optical loss. We can conclude that the worst-case insertion loss for Diamond configuration is higher than both Clements and Reck due to using a higher number of MZIs to perform the same matrix-vector multiplication. Note that SOA gain is not considered in these simulations focusing on the optical loss and crosstalk in OIUs. In addition, observe that the insertion loss for the Clements configuration is almost similar on all the outputs due to its more symmetric mesh configuration compared to Reck and Diamond.
\begin{figure*}[t]
\centering
\includegraphics[width=7.2 in]{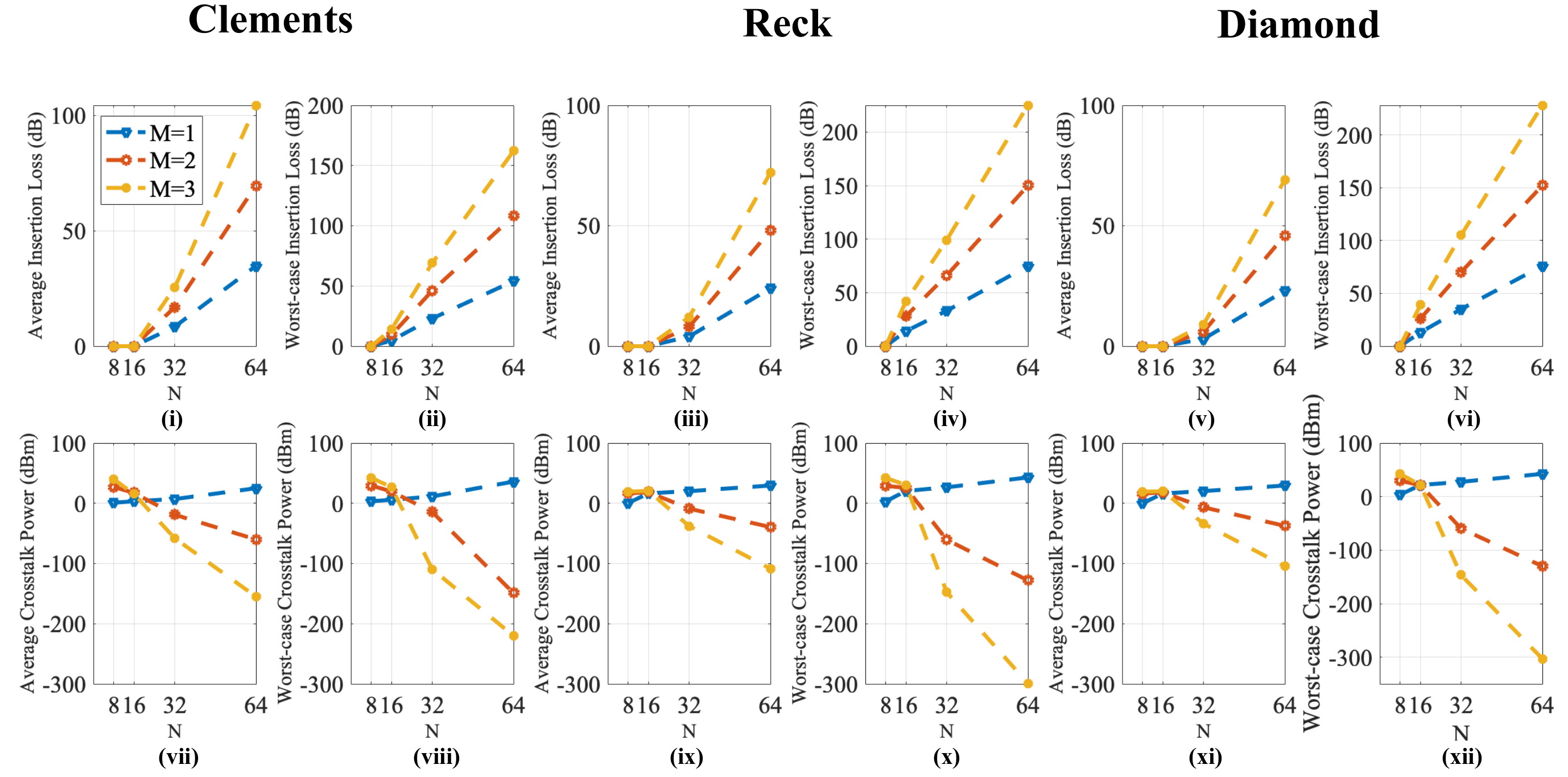}
  \caption{ The average and the worst-case insertion loss and coherent crosstalk power in coherent SP-NNs with different OIU mesh configurations, based on the network in Fig.~\ref{Fig:: MZI-Structure}(b) and with different numbers of inputs ($N$) and layers ($M$). The optical input power at layer one is 0~dBm. 
  }\label{Fig:: Laser_penalty}
  \vspace{-0.15in}
\end{figure*}
We also analyze optical crosstalk noise power for a single layer OIU with $N=$~8 across different mesh configurations. We used 1000 random weight matrices to perform statistical analysis of accumulated coherent crosstalk noise at each output in the OIU with Clements, Reck, and Diamond configurations. In each iteration, a random optical phase ($\rho$), where $0\leq\rho\leq2\pi$, is assigned to the crosstalk noise signal from each MZI in the OIU structure to emulate the crosstalk noise signal phase and behavior throughout the network. Note that the crosstalk signal from each MZI will interact with each other and the victim signal at the outputs of the OIUs. This approach is acceptable when optical signals traverse a large network of devices (e.g., in OIUs), and hence experience random phase shifts. The results related to the crosstalk power including the insertion loss are shown in Fig.~\ref{Fig:: Single_Layer_xt}. Note that 0~dBm input optical power at the input of the OIUs has been considered in these simulations. Observe that the average crosstalk power for the Clements, Reck, and Diamond meshes can be as high as -24.3, -22.1, and -21.41~dBm, respectively. As for the worst-case crosstalk power, the three mesh configurations of Clements, Reck, and Diamond exhibit -6.3, -5.1, and -5.2~dBm, respectively. 
Observe that the crosstalk power is slightly lower on the ports with higher insertion loss in Reck and Diamond configurations, which is consistent with the results in Fig.~\ref{Fig:: Single_Layer_loss}.

\subsection{Optical Loss and Crosstalk in SP-NNs}\label{worst_case_loss_XT}
The optical loss and crosstalk power can be analyzed in an SP-NN comprising multiple hidden layers and with different dimensions and configurations
using the models developed in \eqref{eq::layer_loss} and \eqref{eq::layer_xtalk}. We extended the layer-level insertion loss and crosstalk models in \eqref{eq::layer_loss}
and \eqref{eq::layer_xtalk} for the full-network analysis to demonstrate how the crosstalk power and insertion loss impact changes as we scale up SP-NNs. Average- and worst-case insertion loss and crosstalk analysis are performed for 1000 random weight matrices for SP-NNs with different dimensions ($N=$~8, 16, 32, and 64) and numbers of hidden layers ($M=$~1, 2, and 3) while considering OGUs including SOAs with a gain up to 17~dB (see Table \ref{Table:: Table1}). 
The worst-case and average-case insertion loss and crosstalk power 
at the output of MZI-based SP-NNs with different dimensions and configurations are depicted in Fig.~\ref{Fig:: Laser_penalty}. Note that the insertion loss of the optoelectronic NAU is included in the results. Moreover, for the average case, the mean of the optical loss and crosstalk power over the outputs of the SP-NNs of different dimensions and configurations has been calculated. Consequently, to calculate the worst-case loss and  crosstalk power, the maximum of the aforementioned parameters among all of the SP-NNs' outputs has been reported.

\begin{figure*}[t]
\centering
\includegraphics[width=7.17 in]{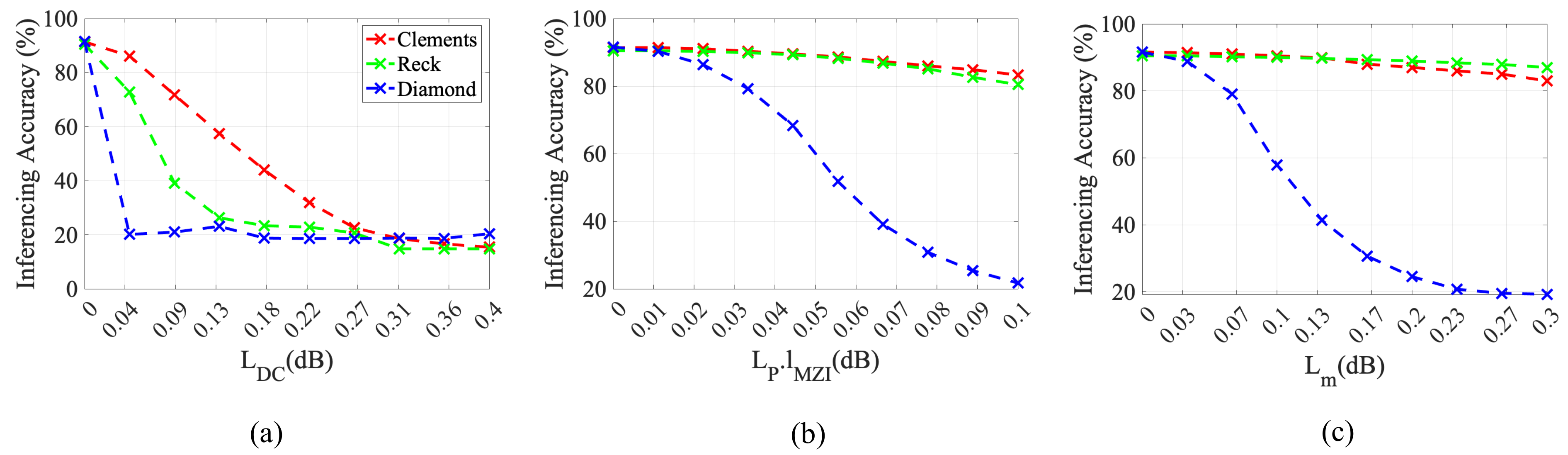}
  \caption{SP-NN inferencing accuracy in the presence of (a) DC insertion loss ($L_{DC}$), (b) propagation loss in the MZI ($L_{MZI}\cdot L_p$), and (c) absorption loss through phase shifter metal
planes ($L_m$).}\label{Fig:: standalone_loss}
  \vspace{-0.15in}
\end{figure*}

As can be seen from Fig. \ref{Fig:: Laser_penalty}, the average (see Fig.~\ref{Fig:: Laser_penalty} (i), (iii), and (v),) and worst-case (see Fig.\ref{Fig:: Laser_penalty} (ii), (iv), and (vi)) insertion loss for SP-NNs increase drastically as $M$ and $N$ increase. As for the Reck and Diamond configurations, the average insertion loss is almost the same and lower than the Clements configuration. The reason for this reduction in loss is the structural differences among the Clements, Reck, and Diamond mesh configurations. As can be seen from Fig.~\ref{Fig:: Architectures}, for Reck and Diamond, the longest and the shortest optical path from the input to the output are not as symmetric as those in Clements. For Reck and Diamond, the shortest path from the input to the output can include only one MZI, while for Clements, it includes $\frac{N}{2}$ MZIs. As for the longest path between the input and the output, Reck and Diamond cross a higher number of MZIs (as high as $N-1$), which is higher than Clements with $\frac{N}{2}$ due to its symmetry in all directions. This also leads to significantly higher worst-case insertion loss for Reck and Diamond (see \ref{Fig:: Laser_penalty} (ii), (iv), and (vi)). For example, the worst-case insertion loss for a single 64$\times$64 layer OIU unit with Reck (see Fig.~\ref{Fig:: Laser_penalty}(iv))) and Diamond (see Fig.~\ref{Fig:: Laser_penalty}(vi) can be as high as 76 dB which is significantly higher than Clements with 54 dB (see Fig.~\ref{Laser_penalty}(ii)). 
Furthermore, the optical loss values scale linearly (see Fig.~\ref{Fig:: Laser_penalty}(i)-(vi)) with the scale of the SP-NNs. Note that although the Diamond mesh has a higher number of MZIs compared to Reck when being used in the OIU units, some of the output ports ($N-2$) are not used during inferencing and are reserved only for the characterization and calibration of the MZIs in the Diamond mesh \cite{shokraneh2020diamond}. This leads to similar average loss and crosstalk performance for SP-NNs with both Reck and Diamond mesh configurations.

Following the same approach, the average- and worst-case crosstalk power is calculated for SP-NNs of different sizes and mesh configurations considering 0~dBm input optical power, as shown in Fig.~\ref{Fig:: Laser_penalty}(vii)-(xii). When $N$ and $M$ increase, the number of MZIs that can generate coherent crosstalk to be accumulated at the output ports increases as well, hence one would expect a higher crosstalk power at the outputs. However, crosstalk signals also experience a higher insertion loss as the network scales up (see Fig.~\ref{Fig:: Laser_penalty}(i)-(vi)). Consequently, the coherent crosstalk power in the output decreases when both $N$ and $M$ increase because of high optical insertion losses as the network scales up. 
For example, when $M=$~1, the average and worst-case coherent crosstalk power increases with $N$ and it can be as high as, respectively, 21.2~dBm and 40~dBm for Clements and 30~dBm and 43~dBm for Reck and Diamond when $N=$~64. We can see that Reck and Diamond show much higher average and worst-case crosstalk power due to their asymmetric mesh structure. Note that using SOAs to compensate for the accumulated insertion loss in the SP-NNs leads to further degradation of the SP-NNs performance because of amplifying  the coherent crosstalk signals at the output of the SP-NN. 


\begin{figure*}[t]
\centering
\includegraphics[width=7.25 in]{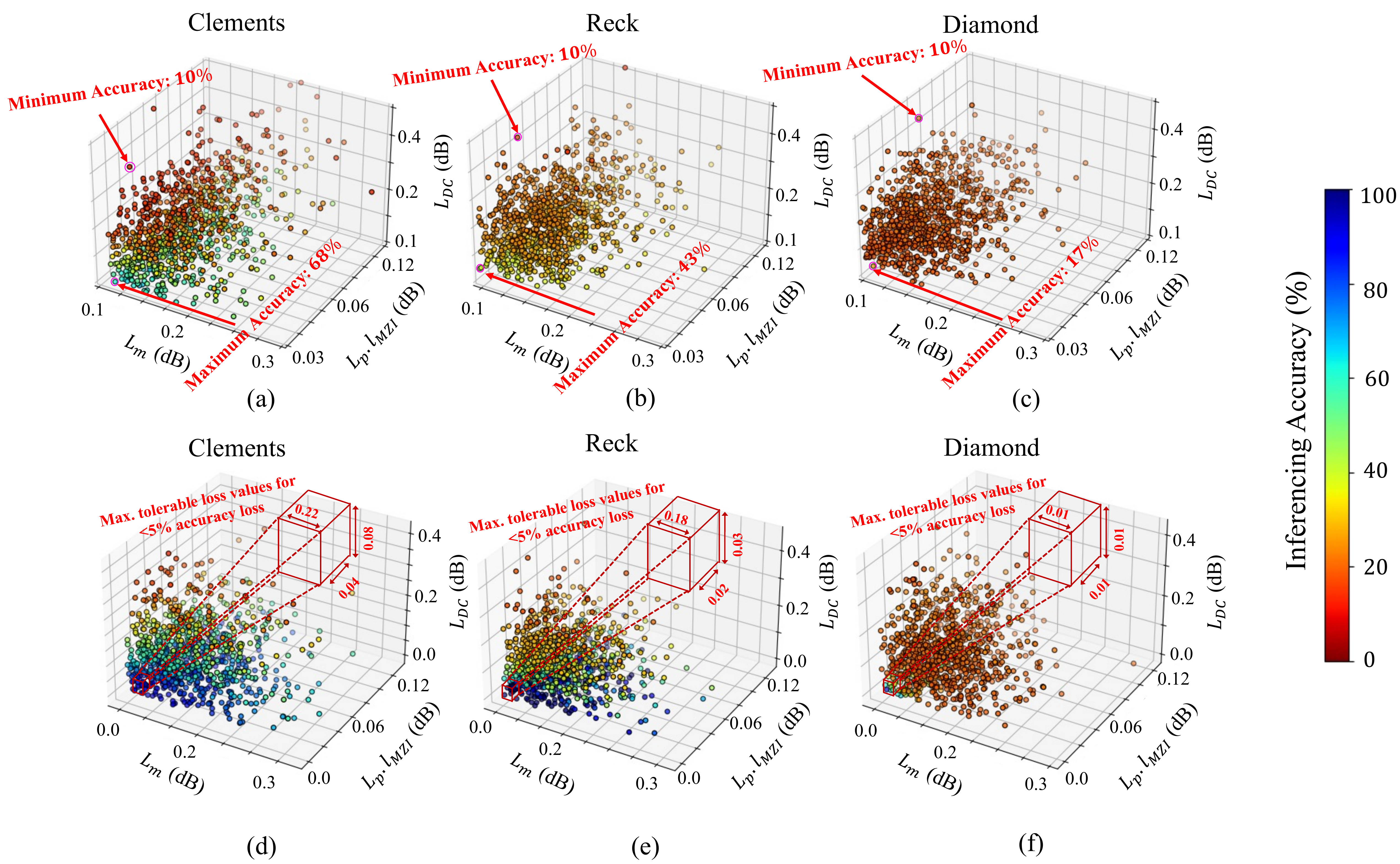}
  \caption{Inferencing accuracy when the loss parameters ($L_{DC}$, $L_{m}$, and $L_p\cdot l_{MZI}$) are simultaneously varied for three different OIU mesh configurations. Two 16$\times$16 hidden layers ($M=3$) have been considered for each case. (a)--(c) show the case where each of the 1000 points in the scatter plot represents an instance of the SP-NN where the $L$'s are sampled from a half-normal distribution with mean, $\mu=$~their minimum expected value and standard deviation, $\sigma$, such that $3\sigma=$~their maximum expected value. (d)--(f) show the case where $L$'s are sampled from a half-normal distribution with mean, $\mu=$~0 and standard deviation, $\sigma$, such that $3\sigma=$~their maximum expected value to show the maximum tolerable $L$ values for each configuration. Note that the effect of the optical crosstalk noise is neglected in these simulations focusing on the effect of optical loss on the SP-NNs inferencing accuracy. }\label{Fig:: simultanous_loss}
  \vspace{-0.15in}
\end{figure*}

\begin{figure*}[t]
\centering
\includegraphics[width=7.17 in]{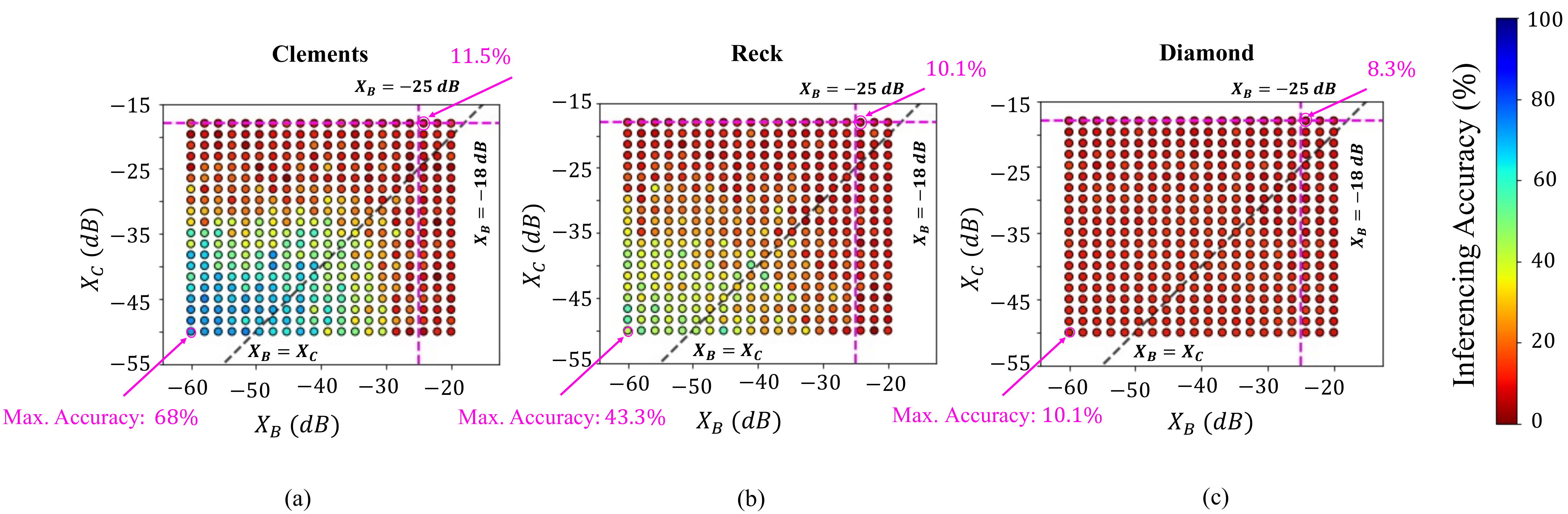}
  \caption{ Inferencing accuracy in the presence of both optical loss and crosstalk noise for different values of $X_B$ and $X_C$ where $X_B\leq X\leq X_C$ (see Section~\ref{Loss_XT_framework}) for different OIU mesh configurations. Two 16$\times$16 hidden layers ($M=3$) have been considered for each case. The minimum expected value for optical loss parameters has been considered in these simulations (See Table \ref{Table:: Table1}). (a) Clements, (b) Reck, and (c) Diamond. The dashed line shows the case where $X_B=X_C$.}\label{Fig:: simultanous_XT}
  \vspace{-0.15in}
\end{figure*}

\subsection{Impact of Optical Loss and Crosstalk Noise on SP-NN Inferencing Accuracy With Ideal NAUs}
To understand the system-level performance degradation in SP-NNs due to the impact of optical loss and crosstalk noise, we consider three case studies of an SP-NN with two hidden layers ($M=$~3) of 16 neurons each ($N=$~16) with Clements, Reck, and Diamond mesh configurations, trained on the MNIST handwritten digit classification task using the same hyperparameters in all the three case studies. Each image in the MNIST dataset is converted to a complex feature vector of length 16 using the shifted fast Fourier transform (shifted-FFT), discussed in \cite{banerjee2021modeling}. The nominal test accuracy after training is 91.5\%, 92.4, and 90.6 for Reck, Clements, and Diamond mesh configurations, respectively. Note that the nominal test accuracy is slightly different for the three case studies due to structural differences in the mesh configurations which leads to different phase settings and distribution over the MZIs in the SP-NNs. To analyze the effect of optical loss and crosstalk noise during inferencing, we integrated the proposed MZI models in \eqref{eq::MZI_trans_mat} and \eqref{eq::MZI_xtalk} into our SP-NN model implementation. Note that in all of the simulations in this section, an ideal \textit{ReLU} nonlinear activation response was considered focusing on the impact of optical loss and crosstalk from OIUs on the SP-NNs inferencing accuracy. Moreover, the OGUs in these simulations are considered to have unity gain during the training due to the normalization of the training and test data.

We consider the expected values of $L_{DC}$, $L_{m}$, and $L_p$ within the ranges 0.1--0.4 dB \cite{Bahadori:16}, 0.1--0.3 dB \cite{ding2016broadband}, and 1--4 dB/cm \cite{Bahadori:16}, respectively. Considering an MZI with a length of $l_{MZI}=$~300~$\mu$m from~\cite{Farhad_4by4}, the propagation loss per MZI ($L_p\cdot l_{MZI}$) is 0.03--0.12~dB. Our analyses showed that the impact of the DC insertion loss ($L_{DC}$) on the SP-NN's inferencing accuracy is significantly higher compared to that of $L_{m}$ and $L_p$ for Clements: the accuracy dropped to $\approx$10\% when only the impact of $L_{DC}$ was considered in the network (see Fig.~\ref{Fig:: standalone_loss}(a)-(c)). Note that the simulation results in Fig.~\ref{Fig:: standalone_loss}(a)  showed the same impact of $L_{DC}$ on inferencing accuracy for SP-NNs with Reck and Diamond mesh configurations when only one source of optical loss was considered at a time. Moreover, the Diamond mesh configuration showed the most susceptibility to the optical insertion loss when one source of optical loss in the MZI was considered at a time (see Fig.~\ref{Fig:: standalone_loss}(a)-(c)).

To understand the simultaneous effect of different optical loss sources (i.e., DCs) on the SP-NN inferencing accuracy, for each case study (i.e., SP-NNs with Clements, Reck, and Diamond mesh configurations), two experiments are defined in which all of the three device-level loss values vary simultaneously from a half-normal distribution. Note that for each experiment, the simulations were repeated 1000 times to avoid loss of generality. The statistical characterization of the two  experiments is as the following: 
\begin{itemize}
    \item (EXPT1) considers a mean of $\mu=$~ minimum expected loss value and standard deviation, $\sigma$, such that $3\sigma=$~their maximum expected loss value. Results for EXPT1 are shown in Fig.~\ref{Fig:: simultanous_loss}(a)--(c).
    \item (EXPT2) considers loss values with a mean of $\mu=$~0 and the same standard deviation as EXPT1. Results for EXPT2 are shown in Fig. \ref{Fig:: simultanous_loss}(d)--(f).
\end{itemize}
As can be seen from EXPT1 results in Fig.~\ref{Fig:: simultanous_loss}(a)--(c), out of the three mesh configurations, Clements shows the most resilience to optical loss. Considering Fig.~\ref{Fig:: simultanous_loss}(a), almost 60\% of the 1000 scenarios show inferencing accuracy less than 25\%. This is much lower compared to the Reck and Diamond which show about 90\% and 100\% of the cases with an accuracy below 25\%, respectively (see Figs.~\ref{Fig:: simultanous_loss}(b) and (c)). Note that in the Clements mesh configuration, only 14\% of 1000 scenarios show inferencing accuracy higher than 50\% and none of the scenarios in Reck and Diamond mesh configurations show accuracy higher than 50\%. Considering the results for EXPT2 shown in Figs.~\ref{Fig:: simultanous_loss}(d)--(c), we can obtain the maximum tolerable device-level optical loss (for each source of optical loss) while considering a threshold for maximum acceptable drop in the inferencing accuracy. In this paper, we consider 5\% accuracy drop in the presence of the simultaneous effect of all sources of optical loss. Accordingly, the Clements mesh configuration can tolerate up to 0.22~dB of metallic loss, 0.04~dB of propagation loss, and 0.08~dB of DC's loss in the MZIs, while these values are lower for the other two configurations (see Figs.~\ref{Fig:: simultanous_loss}(d)--(c)).

Considering \eqref{eq::layer_xtalk}, to understand the impact of optical crosstalk noise on SP-NN inferencing accuracy, we model the crosstalk coefficient $X$ using a linear interpolation between the worst-case (MZI in Cross-bar, $X_C=-$~18~dB) and the best-case (MZI in Bar-state, $X_B=-$~25~dB) crosstalk; see Section~\ref{Loss_XT_framework}. Fig.~\ref{Fig:: simultanous_XT} shows the inferencing accuracy in the three SP-NN mesh configurations in the presence of both optical loss and crosstalk when $X_B\leq X\leq X_C$ and for different $X_B$ and $X_C$ values and with optical losses set to their corresponding minimum expected values. As shown in Fig.~\ref{Fig:: simultanous_XT}, when $X_C=-$18~dB and $X_B=-$25~dB (considered as an example based on the work in \cite{Shoji:10Xtalk}), the accuracy drops to 11.5\% for Clements, 10.1 \% for Reck, and 8.3\% for Diamond. We also found that under expected values of optical crosstalk and loss, the accuracy remains at $\approx$10\% in all three mesh configurations. Note that when $X_{B/C}$ decreases to below -50~dBm (lower-left corner in Figs. \ref{Fig:: simultanous_XT}(a)-(c)), the accuracy saturates at about 68\%  for Clements which is significantly higher than Reck at 43.3\% and Diamond at 10.1\%.  Note that in all of the simulation results presented in this section, the unity gain for the SOAs was considered during the inferencing. The reason for that is that increasing the SOA's gain to compensate for the total insertion loss of the network even further deteriorates the SP-NNs' performance due to amplification of the optical crosstalk noise (see Fig.~\ref{Fig:: Laser_penalty}(vii)-(xii)). In all three case studies presented in this paper, only considering crosstalk noise and neglecting the loss parameters led to an accuracy lower than 25\%. This shows the more critical effect of the crosstalk noise than the loss parameters on SP-NNs' accuracy. Simulation results showed that even a small increase in the SOAs gain ($\approx$5~dB) led to even worse accuracy of below 20\% for all case studies when both optical loss and crosstalk noise were considered because of the amplification of the crosstalk signals at the outputs of the SP-NNs.

The results presented in this section motivate the need for SP-NN design exploration and optimization from the device to the system level to alleviate the impact of optical loss and crosstalk. Moreover, our proposed approach can be used to determine the suitable mesh configuration based on design requirements, and also the maximum tolerable crosstalk power coefficients and component optical losses to guarantee certain inferencing accuracy.
\begin{figure}[t]
\centering
\includegraphics[width=3 in]{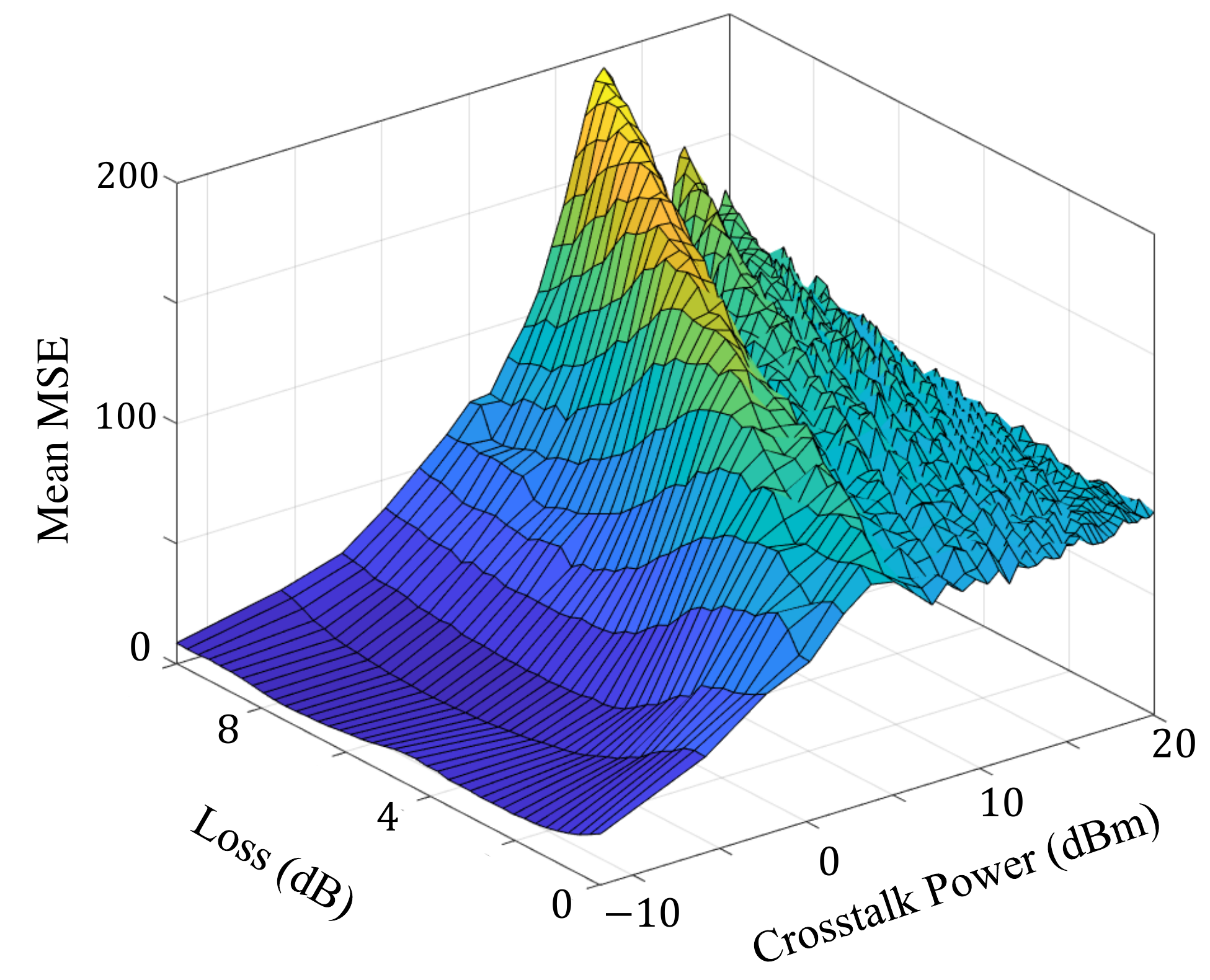}
  \caption{ Average MSE for optoelectronic NAU under the effect of the optical loss and crosstalk noise from OIUs. Each point in the figure has been averaged over 1000 random phases for input crosstalk noise interfering with the victim optical signal.}\label{Fig:: NAU_loss_xt_MSE}
  \vspace{-0.2in}
\end{figure}

\subsection{Impact of Optical Loss and Crosstalk Noise on SP-NN Inferencing Accuracy with Non-Ideal NAUs}
To understand the impact of optical loss and crosstalk noise on SP-NNs' system-level performance when optoelectronic NAUs are used, we use the analytical model in \eqref{eq::NAU_loss_xt} and parameters reported in Table~\ref{table_1} to emulate the \textit{ReLU} as the nonlinear activation function using the NAU depicted in Fig.~\ref{Fig:: NAU}. The mean-square error (MSE) of the NAU's response (compared to the ideal case) with parameters listed in Table \ref{table_1} is used as a metric to understand how optical loss and crosstalk noise from OIUs deteriorates the optoelectronic NAU's performance. 
Note that for each iteration, 1000 random crosstalk noise phases ($\theta_{err}$, see \eqref{eq::NAU_loss_xt}) are considered and the mean of the resulting MSE is reported in Fig.~\ref{Fig:: NAU_loss_xt_MSE}. In this simulation, the PD's sensitivity, dark current, and bandwidth are considered according to Table \ref{Table:: Table1} in the optoelectronic NAU \cite{ding2016broadband}. Moreover, we also consider the optical loss and crosstalk noise related to the input DC and the MZI in the optoelectronic NAU (see Fig. \ref{Fig:: NAU}). Considering the crosstalk noise and loss at the output of the OIU unit, we can see that the MSE for a single optoelectronic NAU used in the SP-NN can be as high as 180 meaning that the actual response of the NAU is significantly deviated from the ideal \textit{ReLU} nonlinear activation response. 
\begin{figure*}[t]
\centering
\includegraphics[width=7.2 in]{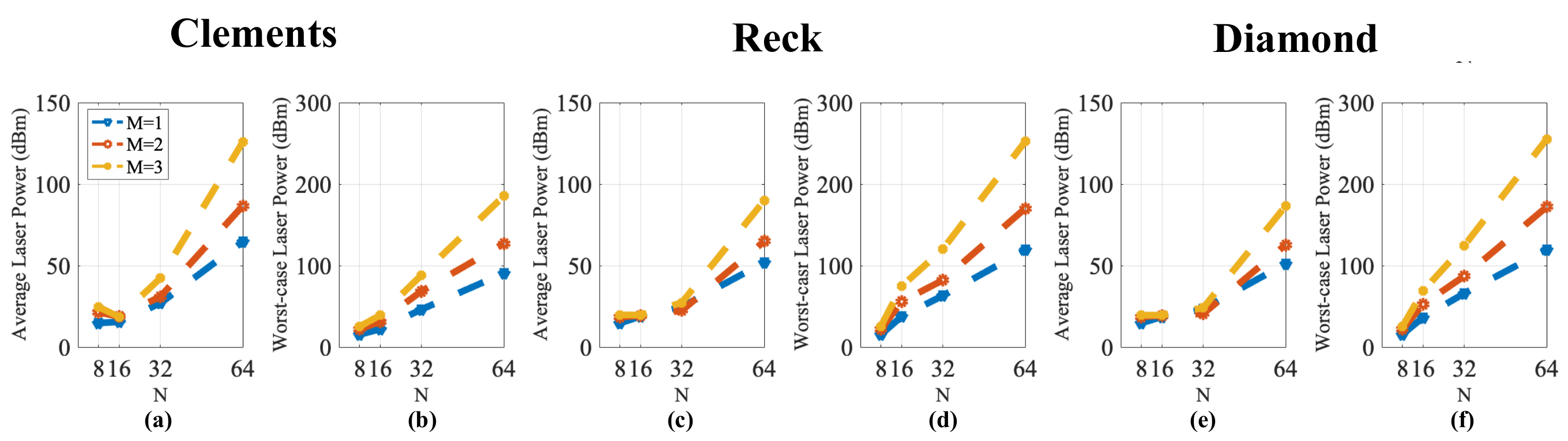}
  \caption{ The optical power penalty in coherent SP-NNs with different OIU mesh configurations, based on the network in Fig.~\ref{Fig:: MZI-Structure}(b) and with different numbers of inputs ($N$) and layers ($M$). 
  }\label{Fig:: optical_pow_penalty}
  \vspace{-0.15in}
\end{figure*}

To analyze the system-level effect of optical loss and crosstalk noise in SP-NNs implemented using optoelectronic NAUs, three different SP-NN networks with 2 hidden layers ($M=$~3) and 16 inputs with Clements, Reck, and Diamond mesh configuration for the OIU and the optoelectronic activation as the NAU (see \eqref{eq::NAU_loss_xt}) were trained on shifted-FFT MNIST handwritten digit dataset using {\fontfamily{qcr}\selectfont
Neuroptica
} with nominal test set accuracy of 90\%, 93.8, and 94.1\% for the Diamond, Reck, and Clements, respectively. Note that also in these simulations, the unity gain has been considered for the OGUs.
We found out that for all three cases when considering the minimum expected values of the device-level loss parameters reported in Table~\ref{table_1} for MZIs in the OIU and optoelectronic NAU, even a negligible crosstalk noise related to the MZIs in the OIUs ($\approx$$-$60~dBm) can lead to a significant drop in the inferencing accuracy to below 15\%, which is drastically lower than the accuracy reported in Fig.~\ref{Fig:: simultanous_XT} when ideal \textit{ReLU} activation function is used (68\% for Clements and 43.3\% for the Reck). Therefore, the SP-NNs which are implemented using optoelectronic NAUs are significantly sensitive to the optical loss and crosstalk noise from OIUs. Note that only considering standalone optical loss and crosstalk from optoelectronic NAUs while neglecting the optical loss and crosstalk from OIUs leads to less than 2\% drop in the SP-NNs inferencing accuracy.

\section{Power Penalty and Scalability Constraint}\label{power_analysis}
Leveraging the results from the previous section, here we analyze the impact of optical loss and crosstalk noise on SP-NN power consumption (i.e., laser power penalty) as well as scalability constraints in SP-NNs.

\subsection{Power Penalty due to Optical Loss and Crosstalk}\label{Laser_penalty}
The optical loss and coherent crosstalk necessitate an increase in the laser power at the input to compensate for optical loss and crosstalk in SP-NNs. We study this power penalty by considering the input optical laser power ($P_{lsr}$) required at the SP-NN input to compensate for the impact of optical loss and coherent crosstalk at the output (see Fig.~\ref{Fig:: MZI-Structure}). For the network output $Y_y$ in a coherent SP-NN, the input optical laser power should satisfy:
\begin{equation*}
    P_{lsr}\geq S_{PD} + IL^y + XP^y(\rho,P_{lsr}).\label{eq::optical_power_penalty}\tag{13}
\end{equation*}
Here, $IL^y$ and $XP^y(\rho,P_{lsr})$ are the insertion loss (in dB) and coherent crosstalk power (in dBm), respectively, at the network output $Y_y$. They can be calculated using \eqref{eq::layer_loss} and \eqref{eq::layer_xtalk}, which include SOA gains in OGUs. Note that the total insertion loss for a signal at output $Y_y$ is determined by both $IL^y$ and the interference between the victim signal and the coherent crosstalk signal (determined by crosstalk signal phase $\rho$) at the same output, where the coherent crosstalk power also depends on $P_{lsr}$. Also, $S_{PD}$ is the sensitivity of the photodetector (in dBm) in electronic or optoelectronic NAUs \cite{PourFard:20}, taken to be $-$11.7~dBm in this paper \cite{9199100PD}. Considering the average and the worst-case insertion loss and crosstalk in Figs.~\ref{Fig:: Laser_penalty}(i)-(xii), Figs.~\ref{Fig:: optical_pow_penalty}(a)-(f) show the average and the worst-case power penalty in coherent SP-NNs of different scales and mesh configurations as $N$ and $M$ increase. Here, the interference between the victim signal and the coherent crosstalk signal at each output is explored statistically and by considering both the average and the worst-case scenarios. On average, the optical power penalty to compensate for both insertion loss and coherent crosstalk is substantially high and easily exceeds 30~dBm for all case studies when $N\geq$32, thereby considerably limiting SP-NN scalability. Note that the worst-case laser power penalty for Diamond and Reck is always significantly higher than Clements (i.e., 254 dBm compared to 188 dBm for $M=3$ and $N=64$).

\subsection{Scalability Constraints due to Optical Loss and Crosstalk}
As it was shown in the previous sections, optical loss and crosstalk significantly limit the scalability of the MZI-based SP-NNs. The work in \cite{liu2022reliability} showed that the matrix-vector multiplications can be done in multiple steps using a single MZI-based OIU. The trained weight matrix can be broken down into multiple sub-matrices and by loading the sub-matrices into memory and repeatedly updating the phase settings of the MZIs in the network, one large matrix-vector multiplication can be carried out using a smaller mesh of OIU to limit the effect of optical loss and crosstalk on the network's performance. Another example of this application is when MZI-based OIUs are used in a crossbar architecture to carry out optical multiplication, which was presented in \cite{giamougiannis2023coherent}.
\begin{figure}[t]
\centering
\includegraphics[width=3.4 in]{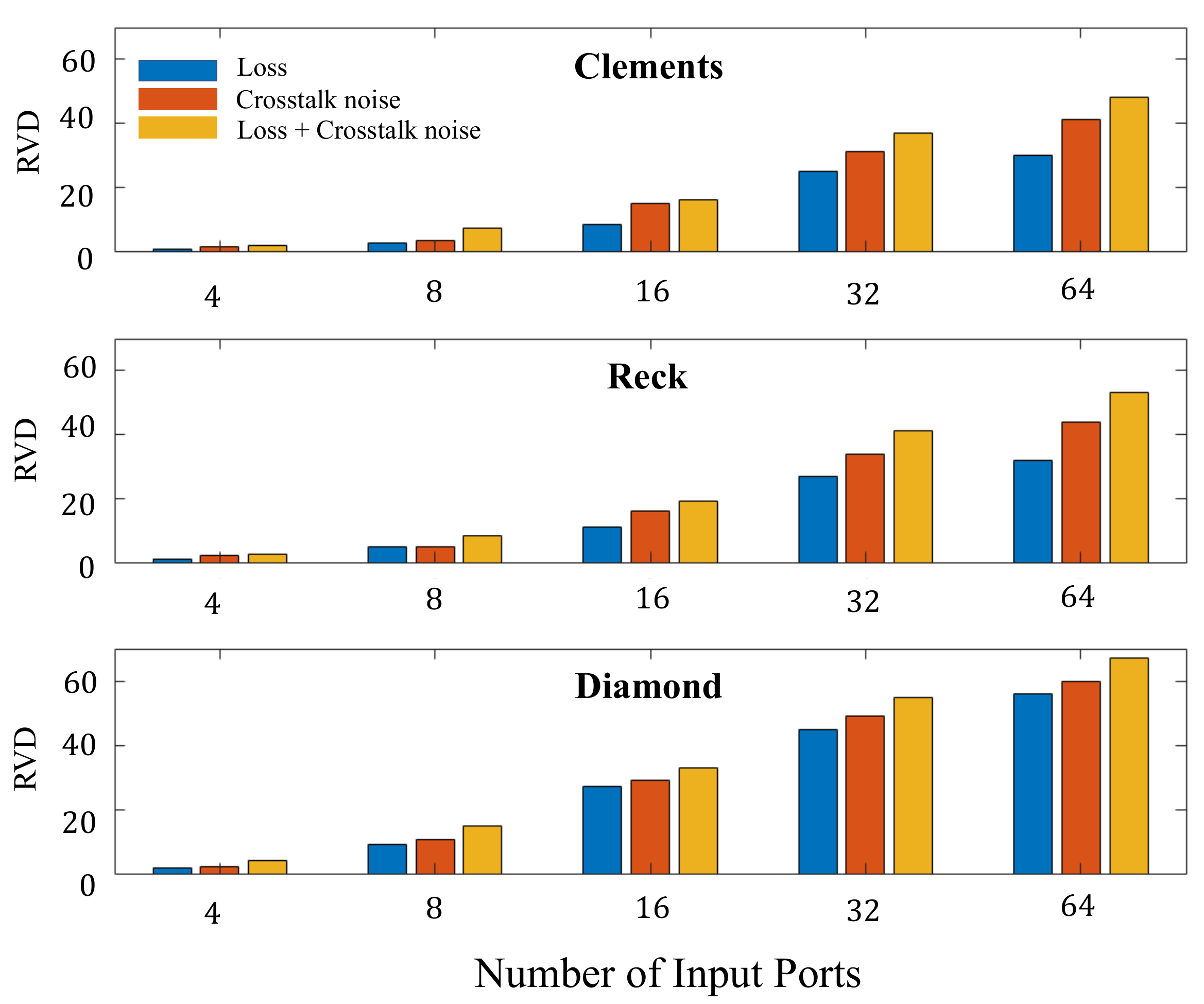}
  \caption{ Average RVD values for 1000 random weight matrices for SP-NNs with different sizes and mesh configurations. Three different scenarios of loss only, crosstalk only, and loss and crosstalk values are considered. }\label{Fig:: scalability}
  \vspace{-0.15in}
\end{figure}

To analyze the scalability constraints due to optical loss and crosstalk noise in SP-NNs for the aforementioned scenarios, we use RVD as a parameter to measure the deviation between an intended transfer matrix and a deviated transfer matrix (due to loss and crosstalk) in an OIU. As a result, this metric can be used to assess how the ideal transfer matrix deviates when optical loss and crosstalk noise are included in OIUs. RVD can be written as:  
\begin{equation}
    RVD(T,\tilde{T}) = \frac{\Sigma_m\Sigma_n |T^{m,n}-\tilde{T}^{m,n}|}{\Sigma_m\Sigma_n |T^{m,n}|}.
    \tag{6}\label{Eq:: RVD}
\end{equation}
In this formulation, $T$ and $\tilde{T}$ are the transfer matrices of the OIU with different mesh configurations with and without the loss and crosstalk noise, respectively. When the RVD is closer to 0, the deviated transfer matrix is closer to the ideal one, hence the inferencing accuracy is higher, as shown by  \cite{mirza2022characterization}. 
Fig. \ref{Fig:: scalability} reports the RVD values for Clements, Reck, and Diamond mesh configurations with different sizes under optical loss and crosstalk reported in Table~\ref{table_1}. For these simulations,  1000 random weight matrices were tested on a single MZI-based OIU with Clements, Reck, and Diamond configuration with sizes of  $N=$ 4, 8, 16, 32, and 64. 
The reason for using 1000 random weight matrices is to avoid loss of generality in our analysis.
As shown by Fig. \ref{Fig:: scalability}, in all cases the impact of optical crosstalk is more critical for scalability than the optical loss (see the results for crosstalk noise in the figure). One reason for this is due to using SOAs to compensate for the total insertion loss of the network that also amplifies the coherent optical crosstalk noise. Moreover, the RVD increases as the network scales up. Observe that the RVD increase is even worse for the Diamond mesh due to using a larger number of MZIs when scaling the network. 

\begin{figure}[t]
\centering
\includegraphics[width=3.6 in]{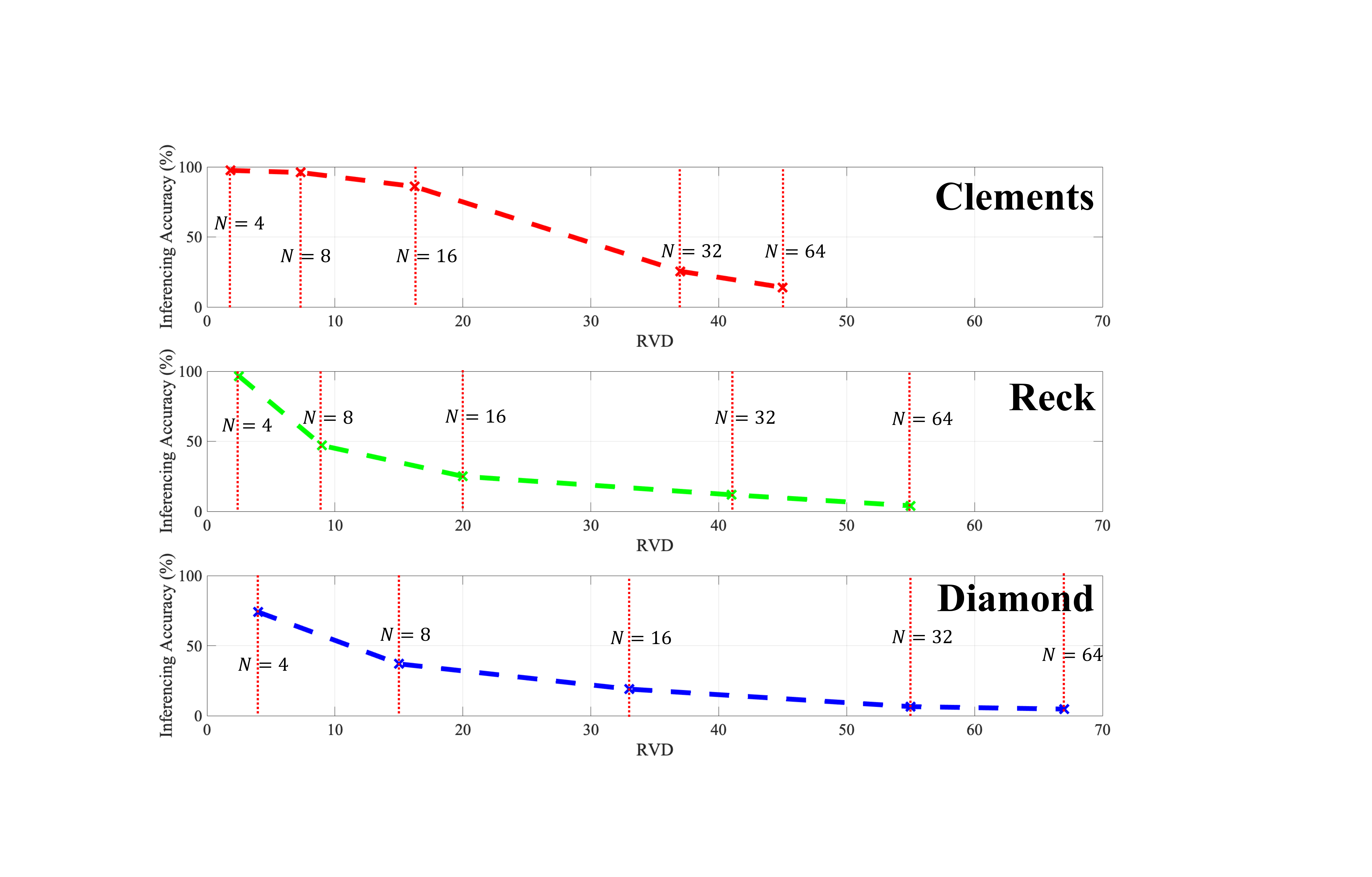}
  \caption{RVD values for  weight matrices for SP-NNs with different sizes and mesh configurations trained on a linearly separable Gaussian dataset against the inferencing accuracy. }\label{Fig:: RVD_acc}
  \vspace{-0.15in}
\end{figure}
To better understand the relationship between RVD and inferencing accuracy of SP-NNs, a single OIU of different mesh configurations with $N=$~4, 8, 16, 32, and 64 followed by an ideal \textit{ReLU} NAU were trained on a linearly separable Gaussian dataset presented in \cite{shokraneh2020diamond}. The test set accuracy is 100\% for all the case studies. Note that the proposed Gaussian dataset is considered because it is simpler than MNIST dataset and requires a single layer only. As a result, we use it as an example to show the relationship between RVD, network accuracy, and the scalability constraints in OIUs across different OIU mesh configurations. 
The RVD values against the inferencing accuracy of a single OIU of different sizes and configurations under the impact of both optical loss and crosstalk noise (using the parameters listed in Table~\ref{table_1}) trained on a linearly separable Gaussian dataset is reported in Fig.~\ref{Fig:: RVD_acc}. Observe that as the RVD increases, the network's accuracy decreases. Moreover, out of three mesh configurations, Clements showed less than 4\% and 15\% drop in the inferencing accuracy when $N=$~8 and $N=$~16, respectively, which is significantly lower than  Reck with 53\% and 75\% and than Diamond with 63\% and 81\% accuracy drops, respectively. The Reck configuration shows about a 4\% drop in accuracy when $N=$~4. Furthermore, the Diamond configuration shows a catastrophic 30\% drop in accuracy even when a deficient number of inputs $N=$~4 is used.

Although the Diamond structure shows the least tolerance to optical loss and crosstalk, the work in \cite{shokraneh2020diamond} suggests that this mesh configuration has the most tolerance to fabrication-process variations compared to Clements \cite{mirza2022characterization}. However, our results presented in this paper show that Diamond mesh cannot be scaled up in the presence of optical loss and crosstalk noise. Note that among the three mesh configurations studied in this paper, Clements architecture showed the highest resilience to optical loss and coherent crosstalk, making it the preferred configuration to implement SP-NNs.

System architects can benefit from the proposed analyses in this paper to better understand the impact of optical loss and crosstalk noise in SP-NNs, and how such an impact change among different SP-NN architecture choices. Also, our analyses can help device designers to better understand device-level performance requirements (e.g., maximum optical loss at the device level) to achieve certain performance and accuracy in SP-NNs. Loss- and crosstalk-aware training of the MZI-based SP-NNs can be considered as a possible solution to alleviate the effect of optical loss and crosstalk in MZI-based photonic computing systems, not studied in this paper.

\section{Conclusion} \label{sec:: 6-Conclusion}

The performance and scalability of SP-NNs are limited by optical loss and crosstalk noise in underlying silicon photonic devices. This paper presents a framework for modeling optical loss and crosstalk noise for SP-NNs of different scales with different mesh configurations. We presented a detailed and comprehensive analysis of average- and worst-case optical loss and crosstalk noise and its corresponding laser power penalty in SP-NNs with three different mesh configurations of Clements, Reck, and Diamond while exploring the drops in inferencing accuracy under different scenarios. In particular, the results showed a significant power penalty to compensate for loss and crosstalk noise
and accuracy loss of at least 84\% for all case studies. 
Additionally, we conducted an extensive analysis of optical loss and crosstalk in optoelectronic NAUs. Moreover, we thoroughly analyzed the scalability limitations of SP-NNs arising from optical loss and crosstalk.
The valuable insights presented in this study can be leveraged by  SP-NN device and system architects to explore and optimize different challenges in the development and evaluation of SP-NNs in the presence of optical loss and crosstalk noise.

\bibliographystyle{IEEEtran}

\bibliography{IEEEabrv,References}

\begin{IEEEbiographynophoto}{Amin Shafiee (GS'22)}
received his BSc degree in Electronics Engineering at Shiraz University, Iran in 2018, and his MS in Integrated Electronics and Optoelectronics at Polytechnique University of Turin (Politecnico di Torino), Italy in 2020. His research interests include Silicon Photonics, Photonic devices, and Photodetectors. He is a graduate research assistant at Colorado State University.
\end{IEEEbiographynophoto}

\begin{IEEEbiographynophoto}{Sanmitra Banerjee}
received the B.Tech. degree from the Indian Institute of Technology, Kharagpur, in 2018, and the M.S. and Ph.D. degrees from Duke University, Durham, NC, in 2021 and 2022, respectively. He is currently a Senior DFX Methodology Engineer at NVIDIA Corporation, Santa Clara, CA. His research interests include machine learning-based DFX techniques, and fault modeling and optimization of emerging AI accelerators under process variations and manufacturing defects.
\end{IEEEbiographynophoto}

\begin{IEEEbiographynophoto}{Krishnendu Chakrabarty}
received the B. Tech. degree from the Indian Institute of Technology, Kharagpur, in 1990, and the M.S.E. and PhD degrees from the University of Michigan, Ann Arbor, in 1992 and 1995, respectively. He is now the Fulton Professor of Microelectronics in the School of Electrical, Computer and Energy Engineering at Arizona State University (ASU). Before moving to ASU, he was the John Cocke Distinguished Professor of Electrical and Computer Engineering (ECE) and Department Chair of ECE at Duke
University. Prof. Chakrabarty is a recipient of the National Science Foundation CAREER award, the Office of Naval Research Young Investigator award, the Humboldt Research Award from the Alexander von Humboldt Foundation, Germany, the IEEE Transactions on CAD Donald O. Pederson Best Paper Award (2015), the IEEE Transactions on VLSI Systems Prize Paper Award (2021), the ACM Transactions on Design Automation of Electronic Systems Best Paper Award (2017), multiple IBM Faculty Awards and HP Labs Open Innovation Research Awards, and over a dozen best paper awards at major conferences.  He is also a recipient of the IEEE Computer Society Technical Achievement Award (2015), the IEEE Circuits and Systems Society Charles A. Desoer Technical Achievement Award (2017), the IEEE Circuits and Systems Society Vitold Belevitch Award (2021), the Semiconductor Research Corporation (SRC) Technical Excellence Award (2018), the SRC Aristotle Award (2022), the IEEE-HKN Asad M. Madni Outstanding Technical Achievement and Excellence Award (2021), and the IEEE Test Technology Technical Council Bob Madge Innovation Award (2018). He is a Research Ambassador of the University of Bremen (Germany) and he was a Hans Fischer Senior Fellow at the Institute for Advanced Study, Technical University of Munich, Germany during 2016-2019. He is a 2018 recipient of the Japan Society for the Promotion of Science (JSPS) Invitational Fellowship in the “Short Term S: Nobel Prize Level” category.Prof. Chakrabarty’s current research projects include: design-for-testability of 3D integrated circuits; AI accelerators; microfluidic biochips; hardware security; AI for healthcare; neuromorphic computing systems. He is a Fellow of ACM, IEEE, and AAAS, and a Golden Core Member of the IEEE Computer Society. He was a Distinguished Visitor of the IEEE Computer Society (2005-2007, 2010-2012), a Distinguished Lecturer of the IEEE Circuits and Systems Society (2006-2007, 2012-2013), and an ACM Distinguished Speaker (2008-2016). Prof. Chakrabarty served as the Editor-in-Chief of IEEE Design \& Test of Computers during 2010-2012, ACM Journal on Emerging Technologies in Computing Systems during 2010-2015, and IEEE Transactions on VLSI Systems during 2015-2018.
\end{IEEEbiographynophoto}

\begin{IEEEbiographynophoto}{Sudeep Pasricha (M'02--SM'13)}
 received his B.E. degree in
Electronics and Communications from Delhi Institute of Technology, India; and his M.S. and
Ph.D. degrees in Computer Science from University of California, Irvine. He is currently a Professor and Chair of Computer Engineering at
Colorado State University, where he is also a
Professor of Computer Science and Systems
Engineering. His research focuses on the design of innovative software algorithms, hardware architectures, and hardware-software co-design techniques for energy-efficient, fault-tolerant, real-time, and secure computing. Prof. Pasricha has received 16 Best Paper Awards and Nominations at various IEEE and ACM conferences, and several other awards for research excellence. He is a Senior Member of the IEEE
and an ACM Distinguished Member.
\end{IEEEbiographynophoto}

\begin{IEEEbiographynophoto}{Mahdi Nikdast (S'10--M'14--SM'19)}
is an Associate Professor and an Endowed Rockwell-Anderson Professor in the
Department of Electrical and Computer Engineering
at the Colorado State University (CSU),
Fort Collins, CO. He received his Ph.D. in Electronic
and Computer Engineering from The Hong
Kong University of Science and Technology
(HKUST), Hong Kong, in 2014. From 2014 to
2017, he was a Postdoctoral fellow at McGill University
and Polytechnique Montreal, Canada. He
is the director of the Electronic-PhotoniC System
Design (ECSyD) Laboratory at CSU. His primary
research goals are focused on the design and development
of photonic systems-on-chip and next-generation data-communication, computing, and sensing systems employing integrated photonics while emphasizing energy efficiency and robustness. Prof. Nikdast and his students have received multiple
Best Paper awards for their work in the area of integrated photonics and
design for manufacturability. He is a senior member of the IEEE.
\end{IEEEbiographynophoto}

\end{document}